\documentclass[reprint,
superscriptaddress,
amsmath,
amssymb,
aps,
twocols
]{revtex4-1} %REVTEX
\usepackage{verbatim}
\usepackage{graphicx}
\usepackage{dcolumn}
\usepackage{bm}
\usepackage{mathtools}

\usepackage{float}

\newcommand\asmap{\ensuremath{\Sigma}}
\usepackage{graphicx}
\usepackage{subcaption}
\usepackage{hyperref}
\usepackage{multirow}
\usepackage{hhline}                         
\usepackage[usenames, dvipsnames]{color}
\usepackage{color}
\usepackage{amssymb, amsfonts}
\usepackage{booktabs}
\usepackage[ruled]{algorithm2e}
\usepackage{algcompatible}
\usepackage{cancel}

\newcolumntype{P}[1]{>{\centering\arraybackslash}p{#1}}

\newcommand{\ie}{i.e., }

\newcommand{\quotes}[1]{``#1''}

\begin{document}

\title{Making sense of complex systems through resolution, relevance, and mapping entropy}

\author{Roi Holtzman}
\affiliation{Department of Physics of Complex Systems, Weizmann Institute of Science, Rehovot, 76100, Israel}
\author{Marco Giulini}
\affiliation{Physics Department, University of Trento, via Sommarive, 14 I-38123 Trento, Italy}
\affiliation{INFN-TIFPA, Trento Institute for Fundamental Physics and Applications, I-38123 Trento, Italy}
\author{Raffaello Potestio}
 \email{raffaello.potestio@unitn.it}
\affiliation{Physics Department, University of Trento, via Sommarive, 14 I-38123 Trento, Italy}
\affiliation{INFN-TIFPA, Trento Institute for Fundamental Physics and Applications, I-38123 Trento, Italy}

\date{\today}

\begin{abstract}
Complex systems are characterised by a tight, nontrivial interplay of their constituents, which gives rise to a multi-scale spectrum of emergent properties. In this scenario, it is practically and conceptually difficult to identify those degrees of freedom that mostly determine the behaviour of the system and separate them from less prominent players. Here, we tackle this problem making use of three measures of statistical information: resolution, relevance, and mapping entropy. We address the links existing among them, taking the moves from the established relation between resolution and relevance and further developing novel connections between resolution and mapping entropy; by these means we can identify, in a quantitative manner, the number and selection of degrees of freedom of the system that preserve the largest information content about the generative process that underlies an empirical dataset. The method, which is implemented in a freely available software, is fully general, as it is shown through the application to three very diverse systems, namely a toy model of independent binary spins, a coarse-grained representation of the financial stock market, and a fully atomistic simulation of a protein.
\end{abstract}

\maketitle

\section{Introduction}

Complex systems challenge our understanding as they resist the reductionist breakdown. A {\it complicated} system can be decomposed into simpler parts and comprehended in terms of their behaviour; on the contrary, a {\it complex} system features a degree of interplay among its constituents that makes its emergent properties impossible to deduce from the study of the irreducible elements it is made of \cite{thurner2018complex,yaneer2019complex,argun2021complex}. In principle, then, these elements should be investigated altogether, simultaneously accounting for their individual behaviour as well as their mutual interactions, correlations, and cooperations.

Nonetheless, a system composed by a large number of degrees of freedom can rarely be understood through a holistic inspection of all of them (it is sufficient to have $\geq 2$ degrees of freedom to have chaotic behaviour \cite{vulpiani2009chaos,strogatz2015chaos}). A substantial decrease of the amount of detail is necessary to attain two goals: on the one hand, the reduction in the sheer number of variables a human mind has to simultaneously cope with; on the other hand, the separation of the relevant information from the irrelevant noise, that is, those properties whose knowledge does not contribute significantly to comprehension. These operations constitute the core business of those methods devoted to {\it dimensionality reduction}.

Many examples of dimensionality reduction algorithms exist \cite{van2009dimensionality,tribello2019using}, such as principal component analysis (PCA), clustering, diffusion maps, intrinsic dimension, and machine learning (ML) approaches. All these provide information about the properties of the system by ``condensing'' the available data about it in to a smaller-sized number of variables that are easier to read, visualise, and interpret. The aforementioned methods are very general in their applicability, and hence the kind of information they provide is similarly general and naturally requires some degree of interpretation to be understood. It is clearly desirable to have methods that are as parameter-free as possible, so as to minimise the amount of antecedent knowledge of the system one has to employ to aptly guide the procedure of simplification; however, ``one-size-fits-all'' approaches are either very hard to conceive or plainly inadequate to tackle the remarkable variety of complex systems that nature offers to those who aim at understanding them. A balance between generality and specificity has then to be found.

A very specific dimensionality reduction strategy is provided by coarse-graining (CG'ing) \cite{noid2013perspective, potestio2014computer, kmiecik2016coarse}, which can make a synthesis between unsupervised feature extraction and a case-specific, easily intelligible analysis of a given system. Originating in the context of critical phenomena \cite{kadanoff1976notes, jose1977renormalization}, coarse-graining was subsequently extended to soft matter modelling \cite{noid2013perspective,giulini2021system}. Here, one aims at constructing simplified representations of molecular systems in which a single super-atom, or bead, is representative of a number of physical atoms; taking advantage of the reduced number of degrees of freedom, the fewer interactions, and the simpler functional form of the latter it is possible to build computationally efficient models that retain the essential qualities of the original system of interest and allow the study of larger molecules for longer times.

Recently, techniques developed in the context of coarse-graining have been employed as instruments not only to {\it model} a system, but also to {\it analyse} a high-resolution model of it, leveraging the fact that the effectiveness of the model largely depends on the appropriate selection of its fundamental constituents. It is in this context that an information-theoretic measure, dubbed mapping entropy \cite{shell2008relative,rudzinski2011coarse,foley2015impact,giulini2020information,giulini2021system,kidder2021energetic}, turned out to be a valuable tool to make sense of a high-resolution model by inspecting lower-resolution representations of it and ranking them according to their mapping entropy value. This quantity, in fact, measures the distance between the reference probability distribution of high-resolution configurations and the one obtained by looking at the system in coarse-grained terms: the lower the mapping entropy, the higher the amount of information retained by a reduced description of the system.

Another approach for studying complex systems is the resolution and relevance framework \cite{marsili2013sampling, haimovici2015criticality, grigolon2016identifying, song2018resolution, cubero2019statistical, cubero2020multiscale, marsili2022quantifying}: here, for a given set of features used to describe the system, the first quantity measures the level of detail this representation provides, while the second quantifies its useful information content. Together, resolution and relevance allow one to pinpoint the level of coarseness that optimally balances data parsimony and informativeness.

In this work we address the problem of identifying novel connections between these distinct measures of information content that have been developed independently in different contexts. We show that these quantities can be employed to differentiate between informative and non-informative features in a sensitive and unsupervised manner, with impactful implications for the comprehension of a large class of complex systems. In particular, we demonstrate that resolution and mapping entropy are strictly connected with one another, and that the combined usage of resolution-relevance first, and mapping entropy later, can constitute a useful data processing pipeline to extract information from empirical data sets.

The paper is organised as follows. In Sec. \ref{sec:background} we present a synthetic overview of the resolution and relevance framework, discuss the derivation and interpretation of mapping entropy, and report novel analytical results on the relation between resolution and mapping entropy. In Sec. \ref{sec:results} we present the results of applying the analysis based on resolution, relevance, and mapping entropy to three distinct systems of increasing complexity. Finally, in Sec. \ref{sec:conclusions} we sum up the results and discuss future perspectives.

\section{Theoretical background}
\label{sec:background}

\subsection{The resolution-relevance framework}
\label{sec:resol-relev-fram-1}

Consider a system composed of $n$ degrees of freedom, e.g. $n$ spins $\sigma_1, \dots, \sigma_n$, whose overall state is specified by the state of each spin. A specific realisation of these spins constitutes an element $\vec{x}$ of an $n$-dimensional vector space. A specific dataset of $L$ configurations, $\{\vec{x}_1,\ \vec{x}_2,\ \cdots\ \vec{x}_L\}$, constitutes the empirical sample that we aim to investigate.

The elements $\vec{x}_i$ of the dataset can be categorised in terms of some labelling $s_i = s(\vec{x}_i)$, where the labels $s$ take values from a discrete set $\mathcal S$ of size $|\mathcal S| = C$. Depending on the classification scheme induced by $s(\vec{x})$, the same label can occur more than once in the same dataset; think, for example, of Ising spin strings classified in terms of their average magnetisation $M = \sum_j\sigma_j$: the value $M = 0$ appears for each string in which half of the spins are up and the other half are down. The number of realisations $\vec{x}_i$ corresponding to the label value $s$ is denoted by $k_s$. The following constraints apply:
\begin{eqnarray}
&&k_s \in \{0, 1, \dots, L\}\\ \nonumber
&&\sum_{s \in \mathcal S} k_s = L,
\end{eqnarray}
meaning that each label can occur a number of time between zero (it never appears) and the size of the empirical dataset (the same label is associated to all data points); furthermore, the occurrences of each label have to sum to the size of the dataset.

The choice of the set of labels induces an empirical probability distribution over the sample given by
\begin{equation}
\label{eq:p_phi}
\hat{p}(s) = \frac{k_s}{L}.
\end{equation}

The Shannon entropy of this distribution
\begin{equation}
\label{eq:hs}
H[s] = - \sum_{ s \in \mathcal S} \frac{k_s}{L} \ln \frac{k_s}{L}
\end{equation}
is termed the \textit{resolution} \cite{haimovici2015criticality}, as it provides a measure of the level of detail employed in the description of the sample. Indeed, a description given by a few labels corresponds to low resolution, as the number of terms in the sum in Eq.~\ref{eq:hs} is small. In contrast, the limiting case where each state has a different label corresponds to a uniform empirical probability $\hat{p}(s) = 1 / L$, leading to the maximal value of the resolution for a sample of $L$ realisations, $H[s] = \ln L$. Intuitively, these two extremes of very gross and very fine descriptions, corresponding to low and high resolution values, do not provide an informative view over the empirical sample; additionally, we observe that the resolution $H[s]$, on average, grows monotonically with the number of labels $C$.

In order to quantify the informativeness of the description given by the classification $s(\vec{x})$, Marsili and coworkers  \cite{marsili2013sampling, haimovici2015criticality, song2018resolution, cubero2019statistical, cubero2020multiscale} proposed to employ the \textit{relevance}: this is given by the Shannon entropy of the distribution of frequencies of labels $s$. Defining $m_k$ the number of labels that have frequency $k$, namely
\begin{equation}
\label{eq:4}
m_k = \sum_{s} \delta_{k, k_s},
\end{equation}
the relevance is then given by
\begin{equation}
\label{eq:hk}
H[k] =  - \sum_{k=1}^L \frac{k m_{k}}{L} \ln \frac{k m_{k}}{L}.
\end{equation}

Note that we omit from the sum those terms for which $m_k=0$, so as to avoid zeros in the logarithm.

The description of an empirical sample in terms of the frequencies of labels $k_s$ provides a {\it minimally sufficient representation} of the sample \cite{cubero2019statistical}. This can be seen by the decomposition of the information content of the sample, the resolution $H_s$, in two parts:
\begin{equation}
  \label{eq:resolution-decomposition}
H[s] = H[k] + H[s|k].
\end{equation}

The first term is the relevance $H[k]$, and the second term is a measure of the noise:
\begin{equation}
\label{eq:noise}
H[s|k] = \sum_{k=1}^L \frac{k m_{k}}{L} \ln m_{k}.
\end{equation}

An intuitive view on this decomposition is the following. The frequency $k_s$ contains information about the label $s$; hence, so does the relevance, which is the entropy of the frequency distribution. Consider now two labels, $s_1, s_2$, having the same frequency $k_{s_1} = k_{s_2}$; in view of the relevance, these labels are equivalent and thus $H[k]$ alone cannot provide any information allowing one to tell them apart. Because of this ambiguity, the term $H[s|k]$ quantifies the degeneracy of the choice of classification scheme $s(\vec{x})$ that produces a specific frequency distribution, and hence it is a measure of noise.

It is now possible to rationalise the intuition for the non-informativeness associated with both extreme values of resolution showcased above. In fact, they both correspond to zero relevance: in particular, when the resolution is zero, all configurations correspond to a single label ${s}$, and thus $k_s = L$ and $m_i = \delta_{i, L}$; analogously, the maximum value of the resolution, $\ln L$, corresponds to a single state per label, namely $k_s = 1 \ \forall \ s$ and thus $m_i = L \delta_{i, 1}$; making use of these values of $m_i$ for the relevance, Eq.~\ref{eq:hk}, gives zero. The non-negativity of the entropy combined with Rolle's theorem implies that the relevance must have a maximum.

Resolution and relevance depend on the specific set of labels $s$ as well as on their number $C$. In general, for small $C$ the resolution is low, and each label has a unique empirical frequency $k_s$ different from that of the other labels. Therefore, knowledge of the frequency implies that of the label, and thus the noise $H[s|k]$ is negligible (see Eq.~\ref{eq:resolution-decomposition}); hence, the relevance is almost equal to the resolution, $H[s] \gtrapprox H[k]$. This linear behaviour is observed in the left part (i.e. low resolution values) of typical resolution-relevance plots, as can be seen in Figs.~\ref{fig:spin}(a-b),~\ref{fig:nasdaq}(a-b) and \ref{fig:cont_model_6d93}(a). Increasing the number of labels $C$, the resolution increases as well. The linear trend $H[k] \sim H[s]$ weakens, until the relevance reaches a maximum and then decreases. Finally, at the highest resolution value $H[s] = \ln L$, the relevance becomes zero.

It is useful to consider the description of the system from the opposite direction, namely going from the maximal resolution and lowering it. This shows that, by reducing the resolution, we actually increase the relevance. The slope of the curve as a function of the resolution, $\mu = \mu(H[s])$, tells us how many bits of relevance we gain by lowering the resolution by one bit. The behaviour of the resolution-relevance curve is extensively discussed by Marsili and coworkers in their analysis of maximally informative samples \cite{song2018resolution, cubero2019statistical}, \ie those sets of realisations of a complex system that maximise the relevance at each value of the resolution. In particular, they identify the threshold point with $\mu = -1$ in these samples as especially interesting, since it provides the optimal trade-off between the two entropies. In the right part of the resolution-relevance plot, the slope $\mu(H[s])$ is generically a decreasing (negative) function of the resolution. Thus reducing the resolution, which corresponds to going from right to left in the resolution-relevance plot, further beyond $\mu= -1$ corresponds to gaining less in relevance than what was lost in resolution.  The point $\mu = -1$ has also been put in relation with a scale-free distribution of frequencies $m_k \sim k^{-2}$, also known as Zipf's law \cite{cubero2019statistical}.

\subsection{Mapping Entropy}
\label{sec:ME}

One of the goals of coarse-graining is to identify a reduced representation, called mapping, of a high-resolution system that retains as much information as possible about it \cite{giulini2021system}. In general, the mapping consists of defining a number $N < n$ of coarse-grained \emph{sites} in terms of a linear combination of the $n$ original degrees of freedom. For the sake of simplicity we here limit ourselves to decimation mappings \cite{kadanoff1976notes, jose1977renormalization, giulini2020information, giulini2021system, menichetti2021journey}, in which a degree of freedom $\sigma_j$ can be either retained or removed from the high-resolution description.

The decimation mapping ${\bf M}$ is defined by the set of indices of the retained degrees of freedom, $j_1, \dots, j_N$, namely:
\begin{equation}
\label{eq:1}
{\bf M}( \sigma_1, \dots, \sigma_n ) = \left( \sigma_{j_1}, \dots, \sigma_{j_N} \right).
\end{equation}

As in the previous section, it is possible to label the different realisations of the system. In this case, we possess a fine-grained label $\vec{x}$, associated to a state of the high-resolution system $(\sigma_1, \dots, \sigma_n)$, and a coarse-grained one $s = s(\vec{x})$, referring to the same configuration, but observed at low-resolution:
\begin{equation}
\label{eq:psi}
s = \left( \sigma_{j_1}, \dots, \sigma_{j_N} \right) \equiv {\bf M} ( \sigma_1, \dots, \sigma_n ).
\end{equation}

Our label $s$ in this case is thus the ($N$-dimensional) string of spins that we retain from the whole. Given this prescription and a coarse-grained mapping ${\bf M}$ (Eq.~\ref{eq:1}), we can now associate the configuration $\vec{x}$ to the corresponding, unique label in the mapped space, $s(\vec{x})$; assuming that the high-resolution states are distributed according to a probability $p(\vec{x})$, we can define a mapped probability distribution in the coarse-grained space $p(s)$, that is the probability of observing the CG label $s$, as:
\begin{equation}
\label{eq:pmacro}
p(s) = \sum_{\vec{x}} p(\vec{x}) \delta_{s(\vec{x}), s}.
\end{equation}

At this point one can introduce the mapping entropy \cite{shell2008relative,rudzinski2011coarse,foley2015impact,giulini2020information,giulini2021system,kidder2021energetic}, which is a Kullback-Leibler divergence measuring the quality of a CG mapping by comparing $p(s)$ to its high-resolution space analogue, $p(\vec{x})$,
\begin{equation}
\label{eq:smap}
S_{map} = \sum_{\vec{x}} \ p(\vec{x}) \ln \left[p(\vec{x}) \frac{\Omega_1 (s(\vec{x}))}{p(s(\vec{x}))} \right] 
\end{equation}
where $\Omega_1 (s)$ is the number of fully detailed, fine-grained configurations $\vec{x}$ mapping onto $s$:
\begin{equation}
\label{eq:omega}
\Omega_1(s)=\sum_{\vec{x}} \ \delta_{s(\vec{x}), s}.
\end{equation}

Specifically, the mapping entropy compares the reference, high-resolution probability, $p(\vec{x})$, against another distribution \cite{rudzinski2011coarse,giulini2020information},
\begin{equation}
\label{eq:pbar}
\overline{p}(\vec{x}) = \frac{p(s(\vec{x}))}{\Omega_1 (s(\vec{x}))},
\end{equation}
which assigns equal probability weight to all the fine-grained configurations that map onto the same CG one. As not all of these configurations are equally probable, these two distributions, $p(\vec{x})$ and $\overline{p}(\vec{x})$, are not equivalent. Ideally, an \quotes{optimal} mapping ${\bf M}$ minimises the impact of the process of dimensionality reduction by aggregating high-resolution configurations with similar probability weight $p(\vec{x})$ inside the same $s$.

The mapping entropy is related to the resolution through (the detailed derivation is provided in Appendix \ref{app:1}):
\begin{eqnarray}
\label{eq:smap_res}
S_{map} &=& \sum_{\vec{x}} p(\vec{x}) \ln \left( p(\vec{x}) \frac{\Omega_1(s(\vec{x}))}{p(s(\vec{x}))} \right)\\ \nonumber
&=& - H[\vec{x}] + H[s] + \sum_s p(s) \ln \Omega_1(s)
\end{eqnarray}
where we have identified $- \sum_{\vec{x}} p(\vec{x}) \ln p(\vec{x})$ as the entropy of the high-resolution representation of the data, $H[\vec{x}]$, and $- \sum_s p(s) \ln p(s)$ as the entropy of the low-resolution representation, hence the resolution $H[s]$. The entropy $H[\vec{x}]$ can be decomposed as follows:
\begin{eqnarray}\label{eq:5}
&&H[\vec{x}] = H[s] + H[\vec{x} | s].
\end{eqnarray}

This equation holds because $s = s(\vec{x})$, that is, the quantity $H[s | \vec{x}] = 0$, since the knowledge of the configuration $\vec{x}$ implies the exact knowledge of the corresponding value of the label $s$. By definition of conditional entropy, the following holds (see Appendix \ref{app:1} for further details on the derivation of this result):
\begin{eqnarray}\label{eq:6}
H[\vec{x} | s] &=& - \sum_{\vec{x}, s} p(\vec{x}, s) \ln \frac{p(\vec{x}, s)}{p(s)}\\ \nonumber
&=& - \sum_{s} p(s) \sum_{\vec{x}} p(\vec{x} | s) \ln p(\vec{x} | s).
\end{eqnarray}

A specific category of classifications $s$ exists that are {\it sufficient representations} \cite{cubero2019statistical}; these are those for which all configurations $\vec{x}$ mapping on a given label $s$ have the same probability, that is:
\begin{eqnarray}\label{eq:7}
&&\forall\ \vec{x}, \vec{x}^\prime\ :\ s(\vec{x}) = s(\vec{x}^\prime),\ \ p(\vec{x}) = p(\vec{x}^\prime).
\end{eqnarray}

Consequently, the conditional probability $p(\vec{x} | s)$ of observing a given data point $\vec{x}$ given the value $s$ of the label is just the inverse of the number of high-resolution configurations mapping on that label:
\begin{eqnarray}\label{eq:8}
&&p(\vec{x} | s) = \frac{\delta_{s(\vec{x}), s}}{\Omega_1(s)}
\end{eqnarray}
where the Kronecker delta is needed to enforce the fact that the conditional probability is different from zero only for those configurations $\vec{x}$ that map onto $s$. Making use of Eq. \ref{eq:8} in Eq. \ref{eq:6} we find that
\begin{eqnarray}\label{eq:9}
H[\vec{x} | s] &=& - \sum_{s} p(s) \sum_{\vec{x}} p(\vec{x} | s) \ln p(\vec{x} | s) \\ \nonumber
&=& \sum_{s} p(s) \ln \Omega_1(s)
\end{eqnarray}
where the last step comes from the definition of $\Omega_1(s)$ given in Eq. \ref{eq:omega}. In conclusion, we have that, if the labelling $s$ is a sufficient representation of the data points, then $H[\vec{x} | s] = \sum_{s} p(s) \ln \Omega_1(s)$. If this is the case, combining Eqs. \ref{eq:smap_res} and \ref{eq:5} with this result we have:
\begin{eqnarray}\label{eq:10}
S_{map} = - H[\vec{x} | s] + \sum_s p(s) \ln \Omega_1(s) = 0.
\end{eqnarray}

This demonstrates that the mapping entropy is the difference between the conditional entropy of the high-resolution data subject to the labelling and the largest value that it can have, which corresponds to $s$ being a sufficient representation. In general, however, resolution and mapping entropy have a nontrivial relation due to the last term in Eq.~\ref{eq:smap_res}.

Changing mapping changes the definition of $s$, and hence the resolution $H[s]$. A mapping that induces a sufficient representation will then have zero mapping entropy; however, the distribution of label frequencies associated to such mapping might not be unique to it, in the sense that other (sufficient) representations might generate the same distribution and, hence, the same relevance. Irrespectively of the mapping entropy being zero, then, the value of the relevance can be smaller or larger depending on the degeneracy of the classifications that produce a given frequency distribution.

\section{Results}
\label{sec:results}

In this work we aim at investigating the behaviour of relevance, resolution, and mapping entropy on distinct systems at varying levels of complexity and abstraction, with the aim of devising a pipeline to process empirical data and extract information out of the dataset. To this end, we concentrated on three different case studies, each of which aims at clarifying specific aspects of the relation among, or possible usages of, these quantities.

First, we made use of a simple toy model to inspect resolution, relevance, and mapping entropy altogether. The system is constituted by a string of non-interacting binary spins; while its properties are trivial to understand once the underlying single-spin probabilities are known, the behaviour of resolution, relevance, and mapping entropy computed on various coarse-grained representations of it is not, as they critically depend on the empirical sample onto which they are computed. This is the ideal situation to (begin to) grasp the essence of these quantities, in that all the non-trivial features that emerge are only marginally due to the complexity of the system itself, and mainly emerging as a consequence of the finiteness of the dataset.

Second, we tackled a real-world case, namely a simplified model of the stock market based on real data. Here, we focus on the relationship between resolution, directly employed as a measure of the detail retained in a given low-detail description of the system, and the mapping entropy, which serves to identify nontrivial correlations within the dataset.

Third, we employed the resolution-relevance framework to reconstruct an empirical probability distribution to be investigated by means of the mapping entropy minimisation method. The latter, in fact, relies on the knowledge of a reference probability distribution of the high-resolution data, against which the low-resolution one is compared. Here we explored the possibility of reconstructing the reference probability from a dataset of protein conformations sampled in a molecular dynamics simulation; to this end, we coarsened the configurational space and identified the reference distribution as the one corresponding to the optimal resolution-relevance threshold ($\mu \sim - 1$).

In the following sections, the results obtained in each of these three systems are presented and discussed.

\subsection{Discrete, non-interacting case: a simple spin system}\label{sec:spin}

\begin{figure}[ht]
    \centering
    \includegraphics[width=\columnwidth]{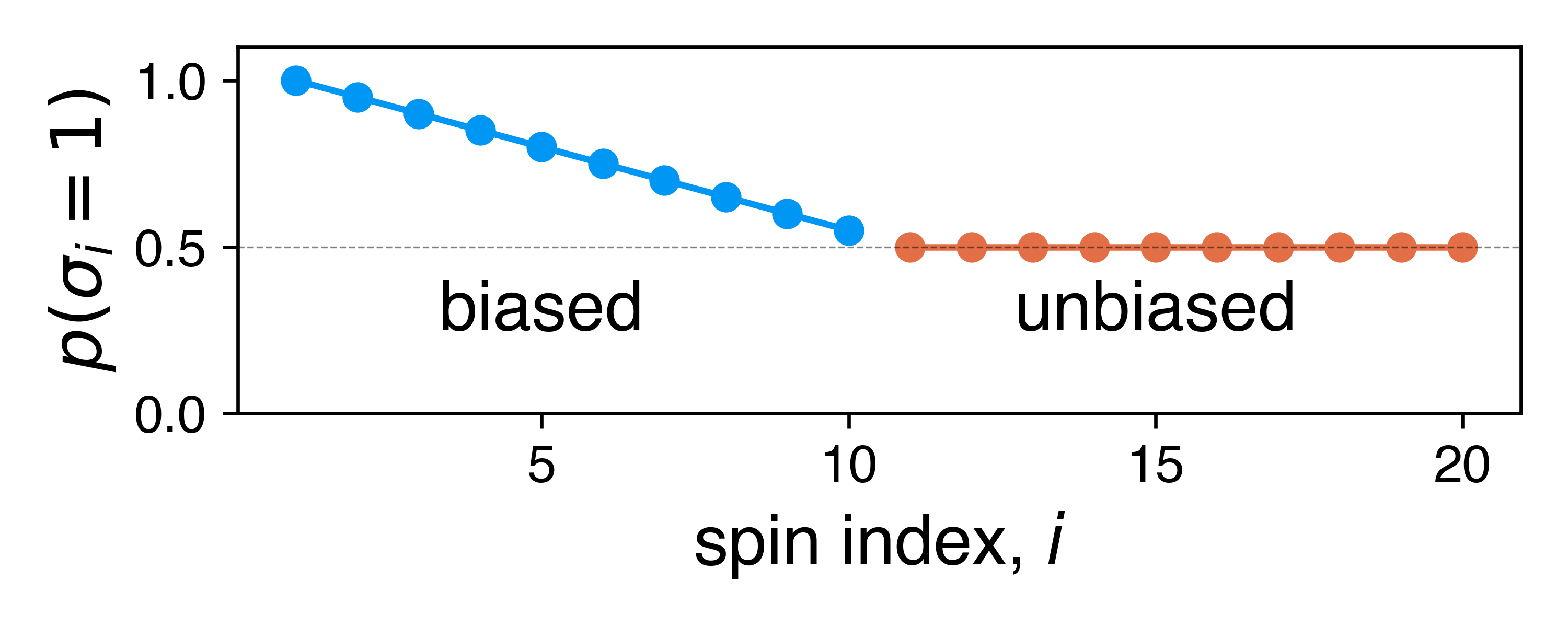}
    \caption{Probability of sampling the ``up'' configuration of each spin. The first 10 spins are biased to a varying degree, whereas the last 10 spins are all unbiased.}
    \label{fig:spin_probs}
\end{figure}

The first model system is composed of $n=20$ non-interacting spins, each characterised by its probability to be in the ``up'' state. These spins are partitioned into two subsets of biased and unbiased spins. The first $10$ spins are {\it biased} in a linear descending order according to $p^A_i (\sigma_i = 1) = 1 - (i-1)/20$ for $1 \le i \le 10$, while the last $10$ spins are {\it unbiased}, namely $p^A_{i} (\sigma_i = 1) = 0.5$ for $11 \le i \le 20$, see Fig.~\ref{fig:spin_probs}.

The number of states of the system is, in principle, $2^{20} \approx 6 \times 10^6$. However, not all of these are realisable as the first spin, $\sigma_1$, has zero probability to be in the ``down'' state. To study this system, we generated a sample of $L=10^5$ states given by $\left\{ \vec{\sigma}^j \right\}_{j=1}^L$. The sample provides an empirical probability distribution of system configurations:
\begin{equation}
\label{eq:pmacro}
\hat{p}(\vec{\sigma}) = \frac{1}{L}\sum^{L}_{j=1} \delta(\vec{\sigma}^{j} - \vec{\sigma}) = \frac{k_{\vec{\sigma}}}{L}.
\end{equation}

In the limit of an infinite sample, the empirical distribution $\hat{p}(\vec{\sigma})$ coincides with the underlying distribution $p(\vec{\sigma})$, that is:
\begin{equation}
\label{eq:2}
\lim_{L \to \infty} \hat{p}(\vec{\sigma}) =  p(\vec{\sigma}) = \prod_{j=1}^n p_j(\sigma_j),
\end{equation}
where the last equality is due to the independence of spins.

\begin{figure*}
    \centering
    \includegraphics[width=1.0\textwidth]{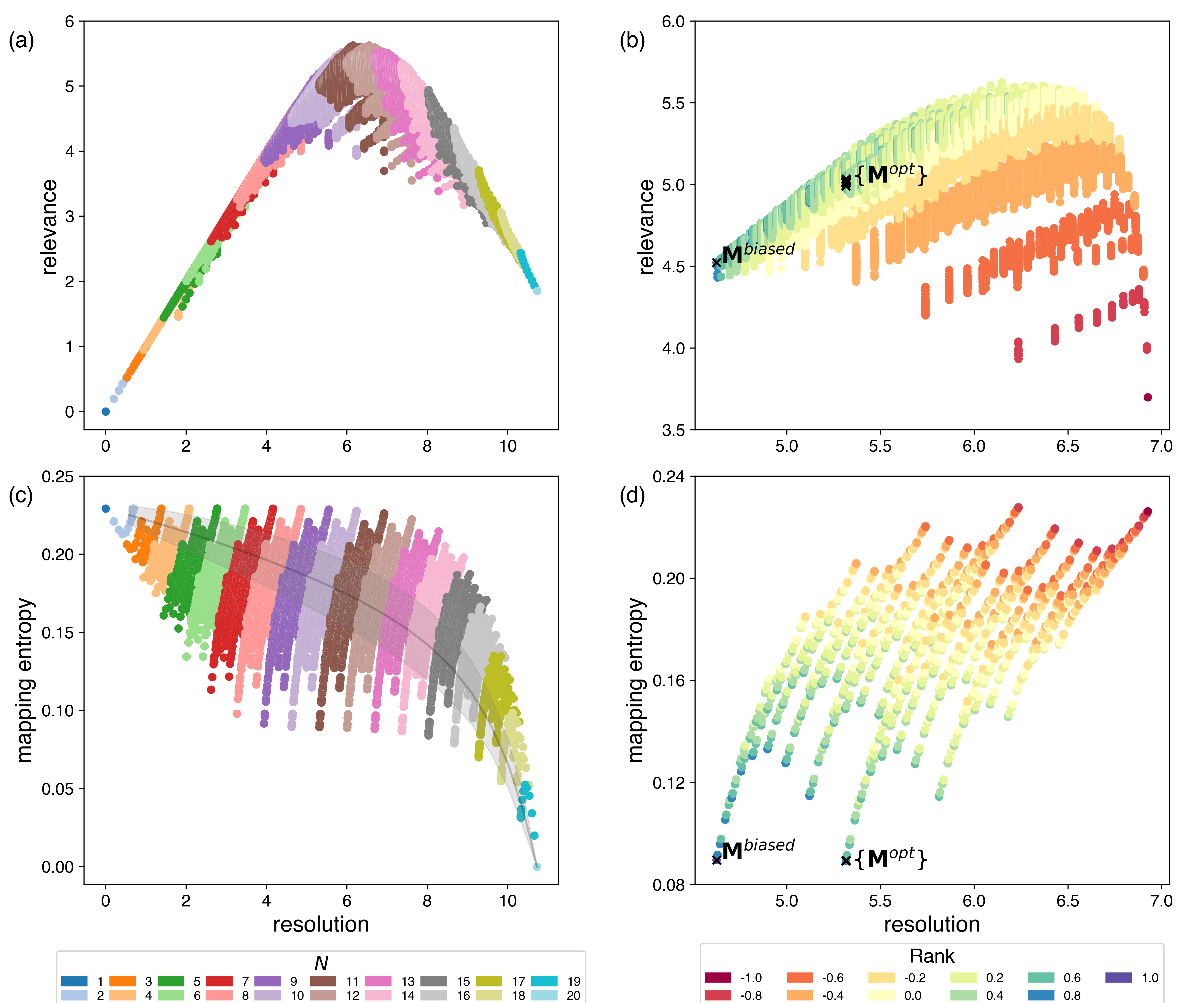}
    \caption{Resolution, relevance and mapping entropy for different coarse-grained representations of the system of non-interacting spins. (a) and (c) show how relevance and mapping entropy vary with increasing resolution. Each data point is depicted according to the number of conserved sites $N$. The gray line in (c) denotes the mean of the mapping entropy for a small range of resolution values, and the shaded region denotes the range of the standard deviation for this range. (b) and (d) report the values of relevance and mapping entropy in the case of $N =10$, respectively. Points are coloured according to their rank, as defined in Eq.~\ref{eq:rank}.  ${\bf M}^{biased}$ marks the mapping in which all the ten biased spins, $\sigma_1, \dots, \sigma_{10}$ are retained, while $\{{\bf M}^{opt}\}$ marks the set of ten mappings displaying the lowest values of mapping entropy. These mappings contain the  spins $\sigma_{2}, \dots, \sigma_{10}, \sigma_{j}$, for $11 \leq j \leq 20$. These are all the biased spins except $\sigma_1$ and  one of the unbiased spins.}
    \label{fig:spin}
\end{figure*}

Let us next discuss  the properties of the coarse-grained representations of this spin system, that is, those selections of $N$ specific spins out of the total $n$. Such a coarse-grained representation is given by a mapping  ${\bf M}: \left\{ 0, 1 \right\}^n \mapsto \left\{ 0, 1 \right\}^N$ which takes the state $\vec{\sigma} = \left( \sigma_1, \dots, \sigma_n \right)$ and returns the CG state $s(\sigma_{1}, \dots, \sigma_{n}) = \left( \sigma_{j_1}, \dots, \sigma_{j_N} \right)$, for some specific choice of $N$ indices $j_1, \dots, j_N$. Each choice of $N$ spins corresponds to another empirical probability of the CG system, $\hat{p}(s)$, which comes about from marginalising over the spins that are not retained. The resolution (Eq.~\ref{eq:hs}) and the relevance (Eq.~\ref{eq:hk}) can be readily calculated: the former directly from the probability $\hat{p}(s)$, the latter through the computation of the frequency distribution, Eq. \ref{eq:4}. To calculate the  mapping entropy one needs to compare the full empirical probability $\hat{p}({\vec{\sigma}})$ with the \quotes{smeared} coarse-grained one, $\overline{p}(\vec{\sigma})$, see Eq.~\ref{eq:pbar}. For each decimation-based CG representation, the corresponding resolution, relevance, and mapping entropy are computed and reported in Fig.~\ref{fig:spin}(a-d).

Specifically, the resolution-relevance values for all possible coarse-grainings of $N=1, \dots, 20$ spins are reported in Fig.~\ref{fig:spin}(a). The first observation we make is that the data follow the expected behaviour in spite of the system being composed of uncorrelated degrees of freedom. The reason for this is that, even if the probability of each spin being ``up'' is independent of the others, the pool of configurations on which resolution and relevance are computed is finite and smaller than the cardinality of possible states ($10^5$ randomly sampled strings vs. $2^{19} \approx 3 \times 10^{6}$ possible ones (recall that the $1$st spin is always ``up''); hence, for about half of the resolution range we are in the {\it under-sampling regime}: because of this, when the resolution is too high, we deal with too few data points to accurately reconstruct the underlying reference probability, and the relevance is lower than the resolution. In the intermediate regime, however, the finiteness of the sample enhances the relevance, and indicates the appropriate resolution level to describe the dataset in a synthetic manner that, nonetheless, allows one to extract nontrivial information about the generative process.

This result is inherently due to the finiteness of the dataset. In fact, if we were to compute resolution and relevance on an exhaustive list of configurations with the exact probability associated to them, the curves would turn out as a band of straight lines, trivially linking resolution and relevance, and with the latter having values below those that are observed in the finite-sampling case (see Sec.~\ref{sec:infinite-sampling} and Fig.~\ref{fig:resrel_inf} in the Appendix).

Another interesting aspect revealed by Fig.~\ref{fig:spin}(a) is the range of resolution and relevance values for different numbers $N$ of retained spins. CG mappings such that $N$ is close to $n$ display little variations in resolution and relevance, while an intermediate coarse-graining is associated with a wide range of values. Figure~\ref{fig:spin}(b) reports the results for the CG representations obtained retaining $N = 10$ sites. Such CG mappings are distributed in a clustered structure that can be captured by introducing a rank for each mapping, which quantifies the balance between biased and unbiased spins. The rank of a single spin $\sigma_j$ is given by
\begin{equation}
  \label{eq:rank-per-spin}
  \tilde{r}(\sigma_j) =
  \begin{cases}
    +1, & \text{if biased:} \ 1 \leq j \leq 10 \\
    -1, & \text{if unbiased:} \ 11 \leq j \leq 20,
  \end{cases}
\end{equation}
and the rank for a CG representation ${\bf M}\left( \sigma_1, \dots, \sigma_n \right) = \left(  \sigma_{j_1}, \dots, \sigma_{j_N}  \right)$ is given by the average of the rank over all retained spins, that is
\begin{equation}
\label{eq:rank}
r({\bf M}) = \frac{1}{N} \sum_{i=1}^N \tilde{r}(\sigma_{j_i}).
\end{equation}

For any choice of $N$, the rank takes a value between $-1$ and $1$, measuring the proportion between biased and unbiased spins in the CG state: when $r({\mathbf{M}}) = 1$ all retained spins are biased; when $r(\mathbf{M}) = -1$ all retained spins are unbiased; when $r({\mathbf{M}}) =0$ there is an equal number of biased and unbiased spins.

Figure~\ref{fig:spin}(b) shows that CG configurations with positive rank provide higher relevance values, whereas negative rank CG configurations have lower relevance and possess higher resolution. Thus, high relevance values correspond to CG mappings that retain more biased spins than unbiased spins, but it is not very sensitive to the rank -- having an equal number of biased and unbiased spins saturates the relevance, i.e. replacing an unbiased spin with a biased one, does not increase the relevance. Therefore, in regards to the question ``which spins are more informative?'' the relevance answers in an ambiguous manner: one should retain just enough biased spins (in this case five), and adding more spins does not change the outcome appreciably.

The reason for this result is a consequence of the marginalised empirical probability of the retained spins. Consider the case of retaining all the unbiased spins: this would provide an empirical sample of labels with a roughly uniform distribution, resulting in a large entropy of the sample and thus a high resolution. As for the relevance, one needs to consider the distribution of frequencies, which in this case would be narrow; as the relevance is the entropy of this distribution, it would correspond to low values. Replacing unbiased spins with biased spins would make the distribution of the sample less uniform, thereby decreasing the resolution. The frequency distribution would become broader, and so the relevance would increase. However, the relevance saturates when we have a rank of zero, i.e. when the number of biased and unbiased spins is equal. This indicates a qualitative feature of the relevance: it thrives when the probabilities of constituents are slightly rather than extremely biased. On the other hand, it increases when retaining constituents with \emph{different} probabilities. Indeed, for a finite sample, the unbiased spins are sampled with finite precision, and therefore, from the empirical point of view, they are slightly biased. Since statistically they are biased in the same manner, retaining too many of them would result in a narrow distribution of frequencies and thus low relevance. However, retaining some of them, already provides enough variability in the frequency distribution to result in high relevance. For further discussion on the differences between infinite and finite samples see the Appendix \ref{sec:infinite-sampling}.

In Figs.~\ref{fig:spin}(c,d) the dependence of the mapping entropy on the resolution is reported. In contrast to the relevance, which tends to zero in the two limiting cases of low and high resolution, the mapping entropy is monotonically decreasing (on average) with the resolution; when all spins are retained, i.e. $N=n$, the smeared probability $\overline{p}(\vec{\sigma})$ (Eq.~\ref{eq:pbar}) is exactly equal to the distribution $p(\vec{\sigma})$ and no coarse-graining is performed; on the other hand, if only one spin is retained, the resulting CG probability is as far as it can be from the full-system probability. For some intermediate values of $N$ it is possible to observe a large range of mapping entropy values, which depend on the specific choice of the CG representation.

Figure~\ref{fig:spin}(d) shows that, for a given $N$, minimal values of the mapping entropy are obtained for high-rank CG configurations, that is, those displaying non-uniform probabilities. A closer look into the minimal values of Fig.~\ref{fig:spin}(d) reveals that the CG mapping (denoted by ${\bf M}^{biased}$ in Fig.~\ref{fig:spin}(b,d)) with maximum rank, $s = \left(\sigma_{1}, \sigma_{2}, \dots, \sigma_{10}\right)$, is not the absolute minimum of the mapping entropy. All of the mappings in which the first spin is replaced by one of the non-biased spins, namely $s= \left( \sigma_{2}, \sigma_{3}, \dots, \sigma_{{10}}, \sigma_{l} \right)$, correspond to lower values of mapping entropy (these are denoted by ${\bf M}^{opt}$ in Fig.~\ref{fig:spin}(b,d)). This is a consequence of the fact that the first spin, having  $p_1 = 1$, is not informative at all (keeping track of its value does not carry any information since it is always ``up''), while each of the non-biased spins provides a minimal advantage due to the finite sample size. In contrast, in the case of fully analytical calculations (which is equivalent to infinite sampling, see Eq.~\ref{eq:2}) the values of the mapping entropy obtained by retaining all the $2, \dots, 10$ spins plus any one of the other eleven spins would be exactly equal.

These considerations allow one to rationalise a feature of Fig.~\ref{fig:spin}(c), namely the fact that the minimum value of the mapping entropy remains approximately constant for a wide range of CG spin numbers, that is, for $N = 9, \dots, 16$. When $N = 9$, the minimum of this quantity is obtained for the CG mapping that retains the spins with indices $2 \leq j \leq 10$, and adding other spins to this representation does not guarantee a substantial decrease in the mapping entropy, which is only obtained when the mapping gets closer to the fully detailed representation (when $N \geq 17$). At the same time, some mappings with $N = 18$ exist, whose associated mapping entropy is higher than the minimum value obtained when $N = 9$: these are coarse-grained representations that do not retain two of the biased spins.

In conclusion of this section, a discrete system whose constituents are completely independent was analysed with the help of resolution, relevance, and mapping entropy. These three quantities shed light on some intrinsic features of the model at hand, thus making them promising candidate analysis tools for more complex systems. In particular, we find that the kind of information highlighted by relevance and mapping entropy is oriented to different goals. The relevance is focused on reconstructing the statistics of the specific empirical sample, and thus it is more ``compression-oriented''. In contrast, the mapping entropy is aimed at marginalising degrees of freedom which do not change the probabilistic description of the sample, and thus it is more ``generation-oriented''. This different sensitivity results in the fact that the mapping entropy favours the biased spins (except $\sigma_1$) over the unbiased spins, while the relevance treats all mappings with zero rank as roughly equal.

\subsection{Discrete interacting case: a model of a financial market}\label{sec:nasdaq}

The second model considered here concerns a simplified description of a financial market, whose constituents are certainly interacting with a functional form that is not only unknown a priori, but also not representative of statistical equilibrium.

Common stock market indices, such as {\sc NASDAQ-100}, {\sc FTSE MIB}, {\sc DAX 30}, are usually defined in terms of the value of the most traded stocks, or the ones with the highest market capitalisation. As an example, the {\sc NASDAQ-100} index considers the largest non-financial companies listed on the Nasdaq stock market \cite{nasdaq100page}. It is well-known \cite{afego2017effects, biktimirov2019asymmetric} that changes in the composition of such indices have an impact on the stock prices, temporarily favouring the stocks that are added to the index.

These indices can be considered as coarse-grained mappings of the high-resolution system, i.e., the full stock market, to a lower number of degrees of freedom. The natural question that arises is the following: are these indices always appropriate to coarse-grain the full market? Can one find a different subset of stocks that brings more information about the high-resolution system?

\begin{figure}
    \includegraphics[width=\columnwidth]{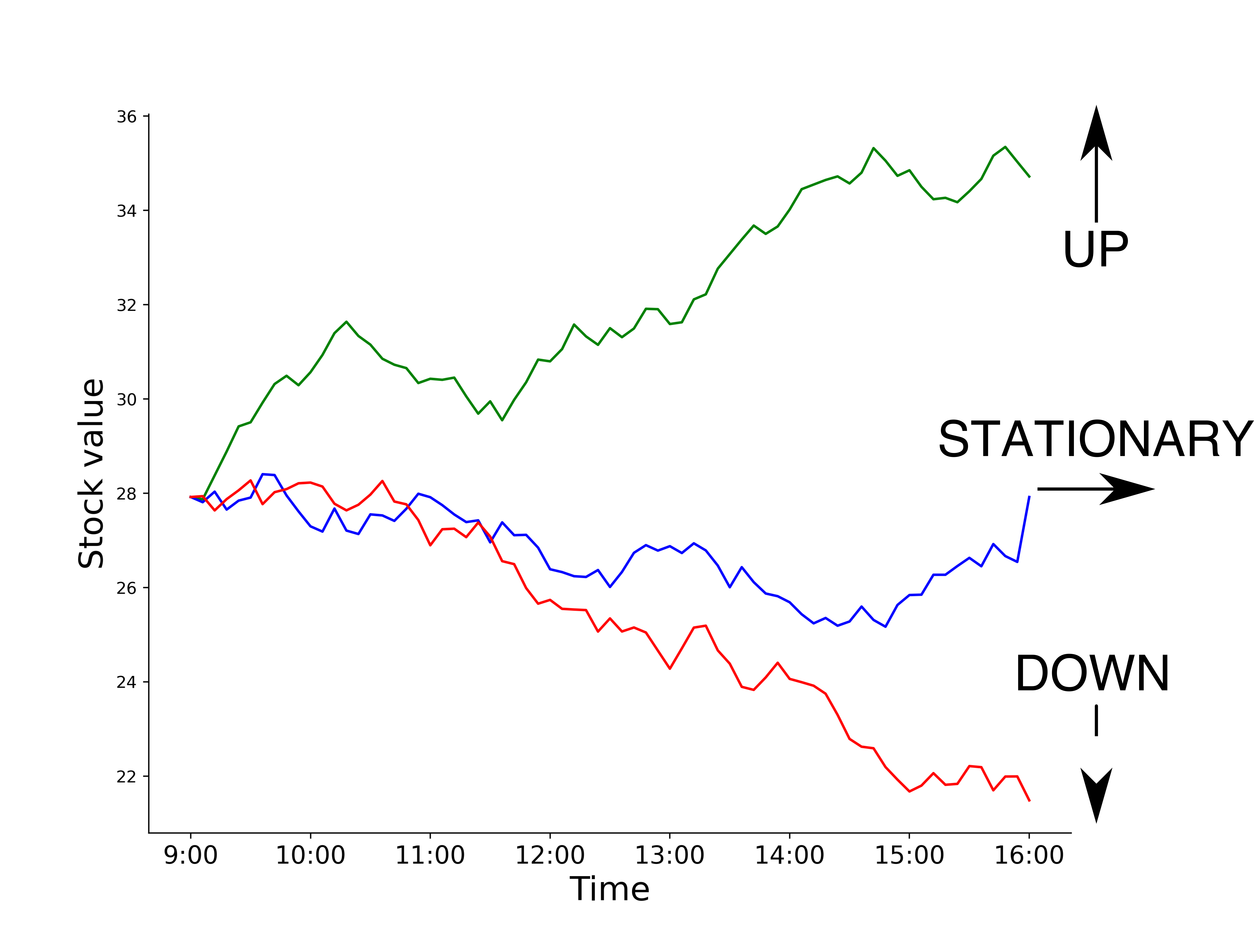}
    \caption{A pictorial representation of the prescription used to build the data set of the discrete model of financial markets illustrated in this section. If the stock value $V$ grows (decreases) during the day with respect to its starting value, that is, if $V_{final} > V_{start}$ ($V_{final} < V_{start}$), a spin \textit{up} (\textit{down}) is assigned to the company for the specific date. If the two values coincide ($V_{final} = V_{start}$), the date is labelled as \textit{stationary} for the considered stock.}
    \label{fig:nasdaq_ex}
\end{figure}

Throughout this section, we consider two ``high-resolution'' systems, namely \textit{m1} and \textit{m2}, defined as the ten (for \textit{m1}) and twelve (for \textit{m2}) stocks with the highest market capitalisation (at the date $1/10/2021$) in the {\sc NASDAQ-100} index, which are described in Tab.~\ref{tab:stocks}. The values of these stocks are investigated over a ten year time window, for a total of $2225$ days of sampling considered. For each day, a stock can assume three discrete values (see Fig.~\ref{fig:nasdaq_ex}), namely $+1$ if the stock value increases during the day, $0$ if it is stationary and $-1$ if it decreases. In this way the full market is mapped to a system of \textit{interacting}, three-states spins with $3^{10}$ ($3^{12}$) available realisations. As in the non-interacting case discussed in Sec.~\ref{sec:spin}, many of these are impossible to observe in a pool of real configurations: imagine for example how unlikely it is that $12$ stocks of this importance are stationary in the same day. Indeed, it is possible to observe only $630$ ($1148$) configurations of the system in the available sampling. As in Sec.~\ref{sec:spin}, we use the set of degrees of freedom as the high-resolution labelling $\vec{x}$, see Eq. \ref{eq:psi}, whose probability $p(\vec{x})$ is defined as the number of times a full-system configuration  $\{{\sigma_1, \dots, \sigma_n}\}$ is observed divided by the number of days (Eq.~\ref{eq:pmacro}).

\begin{table*}
    \begin{tabular}{lcccc}
    \toprule
         Symbol    & Name & \# $\downarrow$ & \# $\rightarrow$ & \# $\uparrow$ \\ \hline
         \addlinespace[0.1cm] {\sc AAPL}  & Apple Inc. Common Stock & 1060 & 5 &1160\\
         {\sc ADBE} & Adobe Inc. Common Stock  & 1017 &   5 & 1203 \\
         {\sc ADI} & Analog Devices, Inc. Common Stock  &1090 & 13 & 1122 \\
         \underline{{\sc CSCO}} & Cisco Systems, Inc. Common Stock & 1022 &  29 & 1174 \\
         {\sc GOOG} & 	Alphabet Inc. Class C Capital Stock  &1069 & 1 & 1155 \\
         {\sc GOOGL} & Alphabet Inc. Class A Common Stock  & 1075 & 2& 1148\\
         {\sc IDXX} & IDEXX Laboratories, Inc. Common Stock  &  977&   8& 1240 \\
         {\sc MSFT} & Microsoft Corporation Common Stock & 1048 &   24& 1153 \\
         {\sc NFLX} & Netflix, Inc. Common Stock  &  1110&    1& 1114 \\
         \underline{{\sc NTES}} &NetEase, Inc. American Depositary Shares&1095&4&1126  \\
	{\sc NVDA} & NVIDIA Corporation Common Stock  & 1078 &  15 & 1132   \\
	{\sc TSLA} & Tesla, Inc. Common Stock  & 1111 &   3 &1111 \\
    \end{tabular}
    \caption{Nasdaq stocks considered in this subsection. \# $\downarrow$, \# $\rightarrow$,and \# $\uparrow$ represent the number of down, stationary and up ``spins'' for each stock during the available sampling time, respectively. {\sc CSCO} and {\sc NTES} are absent in \textit{m1} and are included in \textit{m2}. Data were downloaded using \textit{yfinance} \cite{yfinance}, a python package to download Yahoo! finance data. Companies for which there are no data for all the considered dates were excluded from the dataset.}
    \label{tab:stocks}
\end{table*}

\begin{figure*}
    \centering
    \includegraphics[width=.86\textwidth]{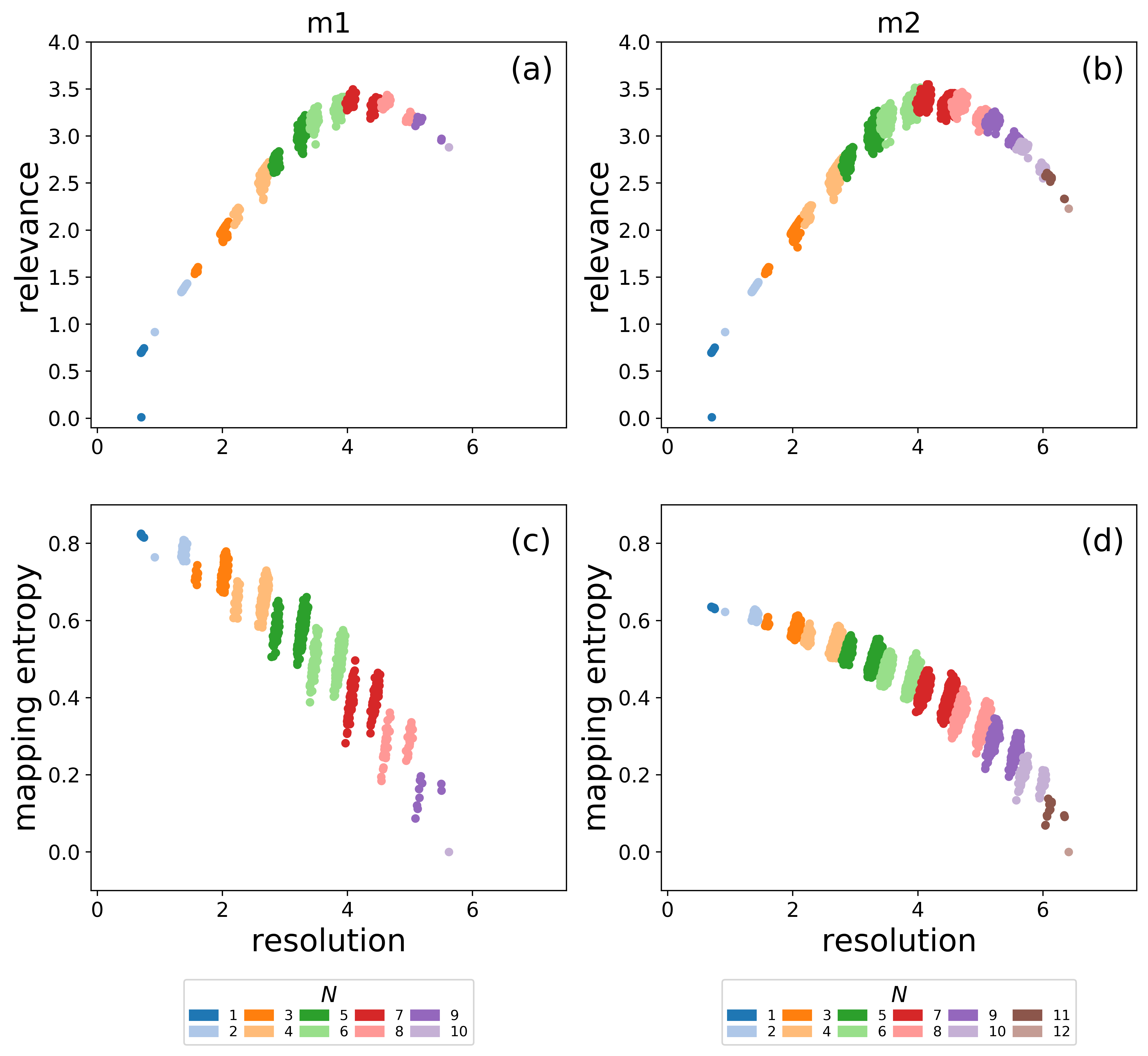}
    \caption{Resolution, relevance ((a-b)) and mapping entropy ((c-d)) for the two models. Mappings in \textit{m2} can reach high values of resolution because adding information (two stocks) allows to define a higher number of high-resolution labels $s$ out of the available sampling. In (a-b) there exists a CG mapping with $N = 1$ possessing a very low value of relevance ($H_k \sim 0$); this is the mapping that retains {\sc TSLA} stock: by chance, the number of spins in the up and down configurations coincide (see Tab.~\ref{tab:stocks}).}
    \label{fig:nasdaq}
\end{figure*}

\begin{figure*}
    \includegraphics[width=\textwidth]{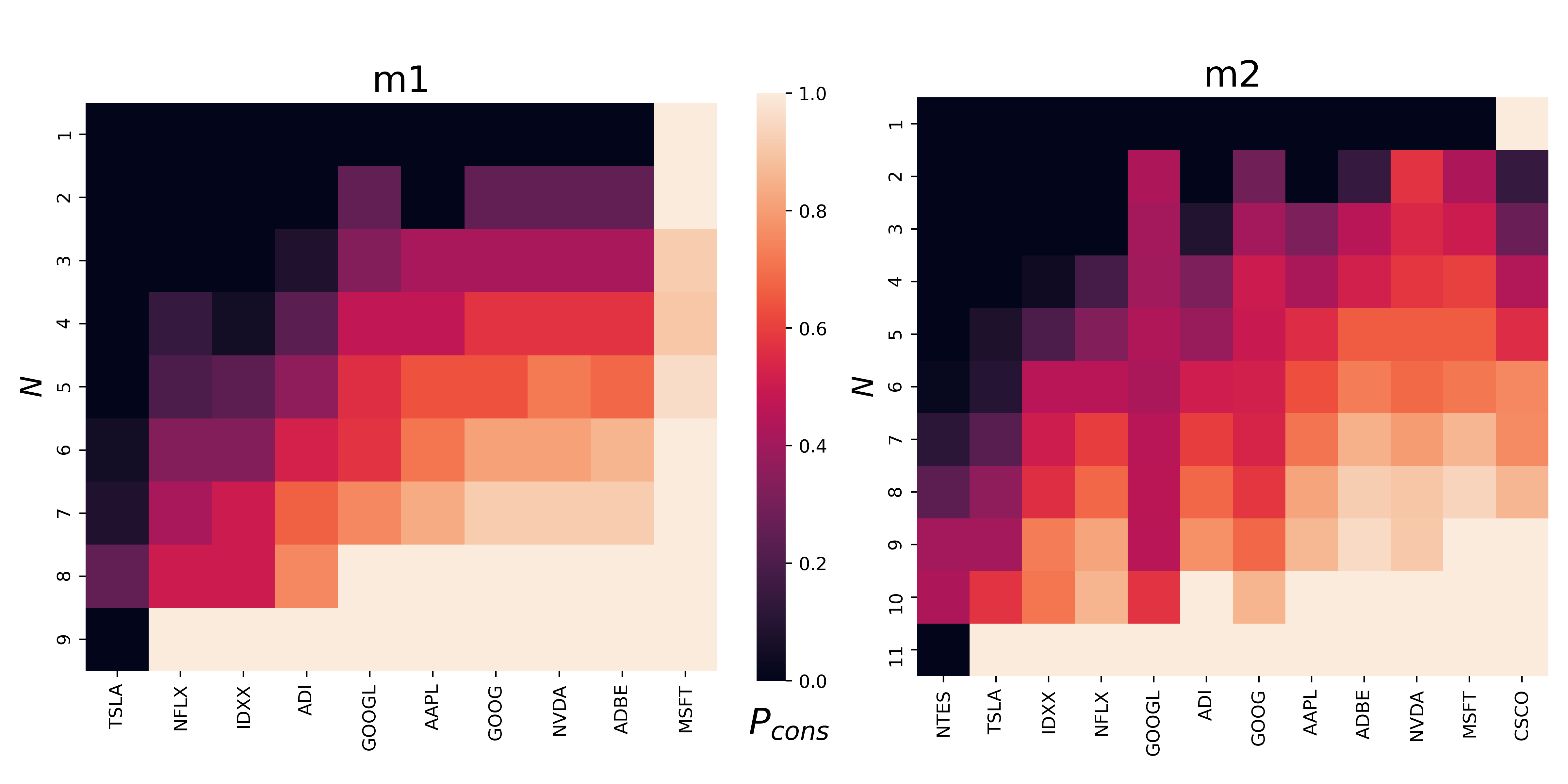}
    \caption{Probability $P_{cons}$ of finding a given stock in a selection that minimises the mapping entropy, for the model \textit{m1} (left panel) and \textit{m2} (right panel). At a given value of $N$, $P_{cons}$ is calculated as the probability of each stock to be present in the 10\% of the mappings with lowest $S_{map}$. Those stocks whose knowledge brings the least information about the overall behaviour of the model market appear in darker colour: the presence of dark bars that extend for a broad range of retained stocks numbers indicates that these specific stocks are consistently identified as little informative.}
    \label{fig:pcons_nasdaq}
\end{figure*}

Next, we analyse the behaviour of resolution, relevance, and mapping entropy for all the $2^9$ ($2^{11}$) CG decimation mappings that can be defined for the two models. The analysis follows Fig.~\ref{fig:nasdaq}, which reports the values of these three quantities for all possible CG mappings, as well as Fig. \ref{fig:pcons_nasdaq}, where we show the probability that a stock is retained in a mapping that minimises the mapping entropy, as a function of the number of retained stocks.

First, looking into the behaviour of the relevance, we observe the expected bell shape, with a linear resolution-relevance trend for $1$ to $4$ retained stocks. This is suggestive of the fact that the model is in the well-sampled regime, and the information content of the dataset is fully captured; for larger numbers of retained sites ($N=5,6,7$), on the contrary, we find a regime where the empirical dataset is noisy, but the coarse representation gathers the largest amount of available information about the underlying statistics. Finally, for $N > 8$, the data are too noisy and the low-resolution representation is not informative.

We then investigate the behaviour of the resolution that is observed in all panels of Fig.~\ref{fig:nasdaq}. For each value of $1<N<n$ there exist two clouds of points separated by a gap in resolution. A direct inspection of the data shows that, at fixed $N$, the lower-resolution clouds of mappings are characterised by a common trait: all these representations retain both {\sc GOOG} and {\sc GOOGL}. As expected, these two stocks are highly interacting and correlated, displaying the same value in the $94.3 \%$ of the selected time window. Therefore, it is reasonable that a mapping containing both Google stocks provides a low-resolution coarse-graining of the system, which is comparable to the resolution of a coarse-grained system with $N-1$ stocks. In Fig.~\ref{fig:nasdaq}(c-d) it is possible to appreciate how the choice of the model influences the average value of mapping entropy of the two clouds. For model \textit{m1} (Fig.~\ref{fig:nasdaq}(c)), the mappings containing both Google stocks (corresponding to the left cloud for each $N$) display an average mapping entropy equal or lower compared to other mappings that contain only one Google stock (corresponding to the right cloud for each $N$). This is not the case for \textit{m2} shown in Fig.~\ref{fig:nasdaq}(d): since two additional stocks are included in \textit{m2}, $p(s)$ is less biased by the presence of Google instances, and the mapping entropy of representations (i.e. mappings) containing both {\sc GOOG} and {\sc GOOGL} is consistently higher than that of the other mappings. Intuitively, one of the two Google stocks possesses a high level of information about the system, but the inclusion of both of them in a coarse-grained description of the full market is redundant.

A further interesting aspect revealed by an inspection of Figs.~\ref{fig:nasdaq}(c-d) and \ref{fig:pcons_nasdaq} is that all the mappings retaining  {\sc GOOG}, {\sc MSFT}, and {\sc NVDA} display a value of mapping entropy lower than the average. In particular it is possible to observe that, in both models, when $3 \leq N \leq n - 1$, the mappings displaying the lowest value of mapping entropy at fixed $N$ always include the combination of these three stocks. The reason behind the high informativeness of these companies can be attributed to their long-time, dominant presence in the stock market.

As for particularly uninformative mappings, that is, those with \textit{high} mapping entropy, it is possible to observe that {\sc TSLA} and {\sc NFLX} (for \textit{m1}) and {\sc TSLA} and {\sc NTES} (for \textit{m2}) appear to be always retained in those representations. In particular, we note that, for \textit{m1} (resp. \textit{m2}), (\emph{i}) when $N = n - 2$ the mapping with lowest mapping entropy is the one that \textit{does not contain} {\sc TSLA} and {\sc NFLX} (resp. {\sc TSLA} and {\sc NTES}); (\emph{ii}) when $2 \leq N \leq n - 2$ the mapping with highest mapping entropy retains {\sc TSLA} and {\sc NFLX} (resp. {\sc TSLA} and {\sc NTES}). A possible explanation for this behaviour can be related to their marginal importance to the market for a vast majority of the sampling time ($10$ years), having experienced an exponential growth only in the latest years. In the case of {\sc TSLA}, the corresponding company operates in a field that is neatly separated from the other stocks reported in Tab. \ref{tab:stocks}.

Lastly, we note that the ``interacting'' system considered in this section does not display the flatness in the mapping entropy minima that was observed in  Fig.~\ref{fig:spin}(c) describing the non-interacting spins system in Sec.~\ref{sec:spin}. In fact, for an interacting system the addition of a new site to an optimal coarse-grained mapping is likely to result in a gain of information about the high-resolution system and, hence, in a decrease of the mapping entropy.

In summary, the information measures under examination, and in particular their joint usage, proved to constitute an informative instrument of analysis of our simple description of a subset of the Nasdaq financial market. Specifically, the resolution-relevance curve was shown to highlight interesting distinct regimes of the low-resolution description, providing a guide in assessing, qualitative and semi-quantitatively, the amount of useful information that a coarse picture of the system can retain; the mapping entropy, on the other hand, allowed us to rationalise the features observed in the resolution and to identify those specific stocks that contributed the most (or the least) to the overall behaviour of the model market. The proposed strategy can thus be generalised to the full stock market with the aim of selecting the most appropriate low-resolution index, identified as the set of stocks with minimal mapping entropy at a fixed degree of coarse-graining $N$, the latter that can be determined with the help of the resolution-relevance curve.

\subsection{Continuous system: a small protein in solution}\label{sec:cont} 

As it has been illustrated in the previous sections, the mapping entropy is a measure of how much information about a reference, high-resolution system (and its configurational probability distribution) can be retrieved or inferred from a low-resolution representation of it. In particular, one computes the Kullback-Leibler divergence between the reference distribution $p(\vec{x})$ and the reconstructed one, $\bar{p}(\vec{x})$, which is obtained from the former assuming that all configurations $\vec{x}$ mapping on the same coarse-grained label $s$ have the same probability; the latter is defined as the average over the group $\mathcal G$ of configurations $\vec{x}_i\ :\ s(\vec{x}_i) = s_{\mathcal G}\ \forall\ i \in\ \mathcal G$.

In this section we address the practical aspect of investigating systems with continuous degrees of freedom, whose reference empirical probability distribution $p(\vec{x})$ has to be determined. The problem lies in the fact that, while systems with discrete degrees of freedom (such as the stock market model) are naturally prone to a histogramming procedure, systems described in terms of continuous variables are not: arbitrarily small discrepancies in the coordinates would make two configurations look different, and whether they really are or not is a matter to be settled before addressing the computation of the mapping entropy.

Here, our objective is to employ the resolution-relevance framework to perform an optimal clustering of the {\it high-resolution} configurations of the system, based on which we determine the reference empirical probability $p(\vec{x})$. This is a key step for the calculation of the mapping entropy: in fact, in specific cases, e.g. molecular systems at thermal equilibrium, the mapping entropy can be computed by means of a cumulant expansion of the Kullback-Leibler divergence that relies on the assumption that the system follows Boltzmann statistics, and hence the underlying probability density of the micro-states is the well-known $\exp(-\beta H)$; this strategy was indeed employed by Giulini and coworkers (see Ref. \cite{giulini2020information} as well as Eq. \ref{eq:smap_cum.1_main} in the Appendix) to identify the representations of least mapping entropy for a set of proteins. This assumption, however, does not hold in general, and it might be the case that one finds themselves with a dataset of configurations defined on a continuum range of values, whose underlying probability density is not known. The computation of mapping entropy in these cases has to rely on the definition based on the Kullback-Leibler divergence, which, in turn, assumes the knowledge of a reference, high-resolution probability density. Hereafter, we show how to obtain such probability distribution for a dataset of configurations defined on the continuum, and demonstrate that the results so obtained are consistent with those derived from the cumulant expansion.

The system under examination here is a small protein in water, whose time evolution is obtained by means of a plain, all-atom molecular dynamics (MD) \cite{alder1959studies, karplus2002molecular} simulation. Specifically, we consider 6D93 \cite{mayorga2020novel}, a mutant of the tamapin protein, a toxin of the Indian red scorpion \cite{pedarzani2002tamapin}. This small protein (230 heavy atoms, 31 amino acids) is simulated in the canonical ensemble at $300 \text{K}$ for $200$ nanoseconds. The Cartesian coordinates of the atoms are saved once every $20$ picoseconds, thus creating a data sample (trajectory) of $L = 10001$ configurations. Details on the GROMACS 2018 \cite{van2005gromacs,abraham2015gromacs} simulation can be found in the Supplementary Material of Ref.~\cite{giulini2020information}.

In this context, the state of the system is encoded in a vector ${\bf r}$  containing the positions of its $n$ constituent atoms. Differently from the discrete model, the distribution of the labels $\vec{x}$ cannot be identified with a simple counting over the states of these $3n$ degrees of freedom, due to their continuous nature.
Hence, the labels $\vec{x}$ have to be defined by lumping several, in principle different configurations ${\bf r}$ of the sample in the same (high resolution) state, thus defining a non-uniform probability $p(\vec{x})$ of observing it.

To this end, we apply the UPGMA clustering algorithm with average linkage \cite{sokal1958statistical} to the fully atomistic, pairwise RMSD matrix between all the elements of the sample:
\begin{equation}\label{eq:rmsd}
    \text{RMSD} ({\bf r} , {\bf r}^\prime) = \sqrt{\frac{1}{n} \sum_{j=1}^{n} \sum_{k=1}^{3}({r}_{3j+k} - \mathcal{RT} {r}^\prime_{3j+k})^{2}},
\end{equation}
where $\mathcal{RT}$ is the roto-translation that superimposes ${\bf r}$ to ${\bf r}^\prime$ according to the Kabsch optimality criterion \cite{kabsch1976solution, kabsch1978discussion}, thus minimising the overall displacement.

\begin{figure}
    \includegraphics[width=\columnwidth]{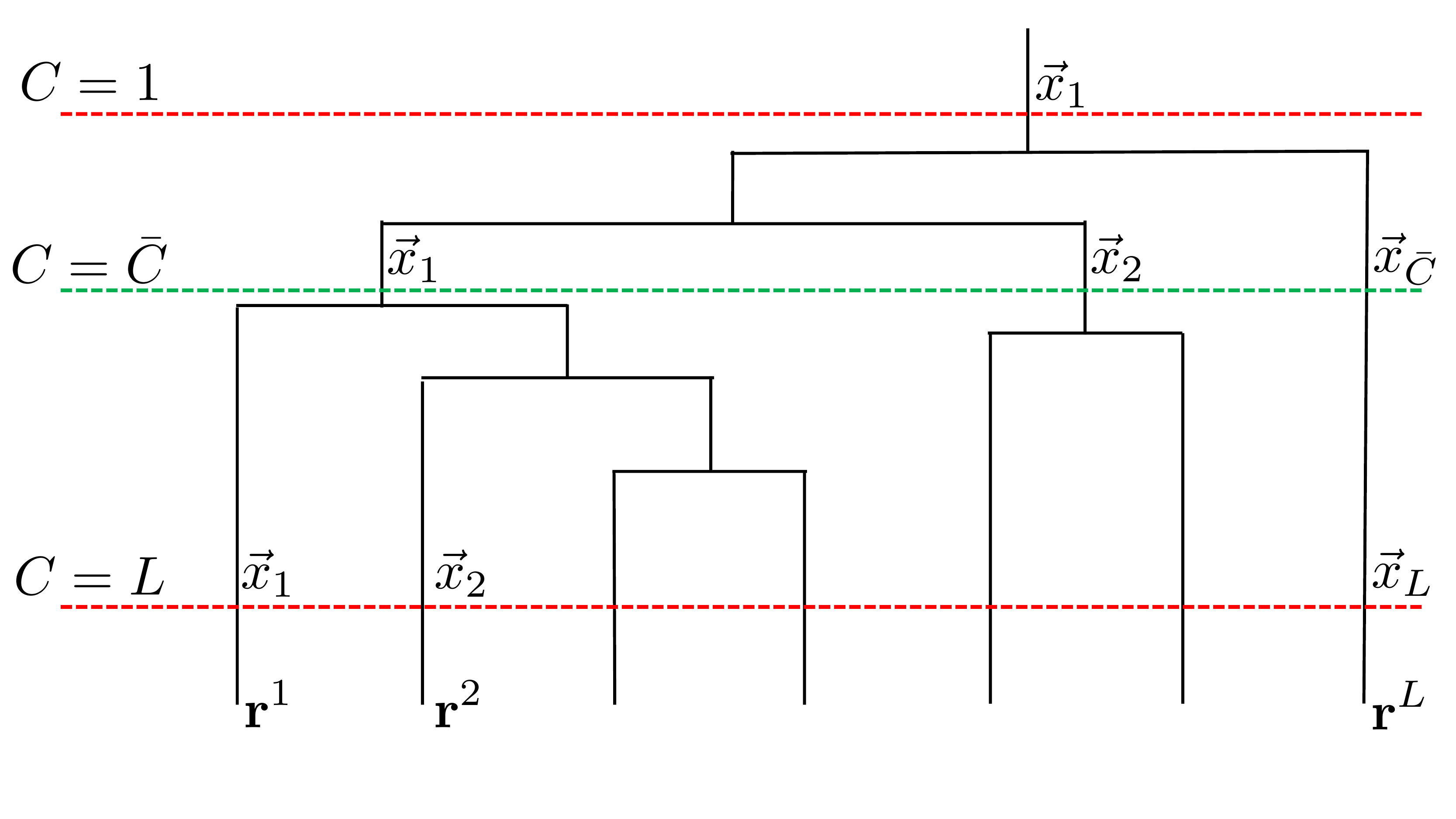}
    \caption{The $L$ realisations of a continuous system can be clustered in a variable number $C$ of labels $\vec{x}$, ranging from $C=1$ to $C=L$. These two discretisations are not informative about the system, as they induce a trivial frequency distribution and, consequently, a uniform probability $p(\vec{x})$ of observing the label $\vec{x}$ over the sample. We identify (see main text) the threshold $\bar{C}$ as the number of labels that separates the regimes of lossless and lossy compression.}
    \label{fig:dendrogram}
\end{figure}

Changing the threshold used to cut the dendrogram (see Fig.~\ref{fig:dendrogram}) resulting from the UPGMA clustering of this matrix,  we can obtain arbitrary values of the number $C$ of fine-grained labels ${\vec{x}}$.  A criterion is thus needed to establish the appropriate value of $C$ that should be used to create this histogram of atomistic structures. A very coarse (resp. detailed) discretisation of the sample corresponds to $C \sim 1$ (resp. $C \sim L$) clusters, as shown in Fig.~\ref{fig:dendrogram}. Both of these choices are not \quotes{relevant} for the comprehension of the system, since they result in a uniform probability $p(\vec{x})$ over the labels.

\begin{figure}
    \includegraphics[width=\columnwidth]{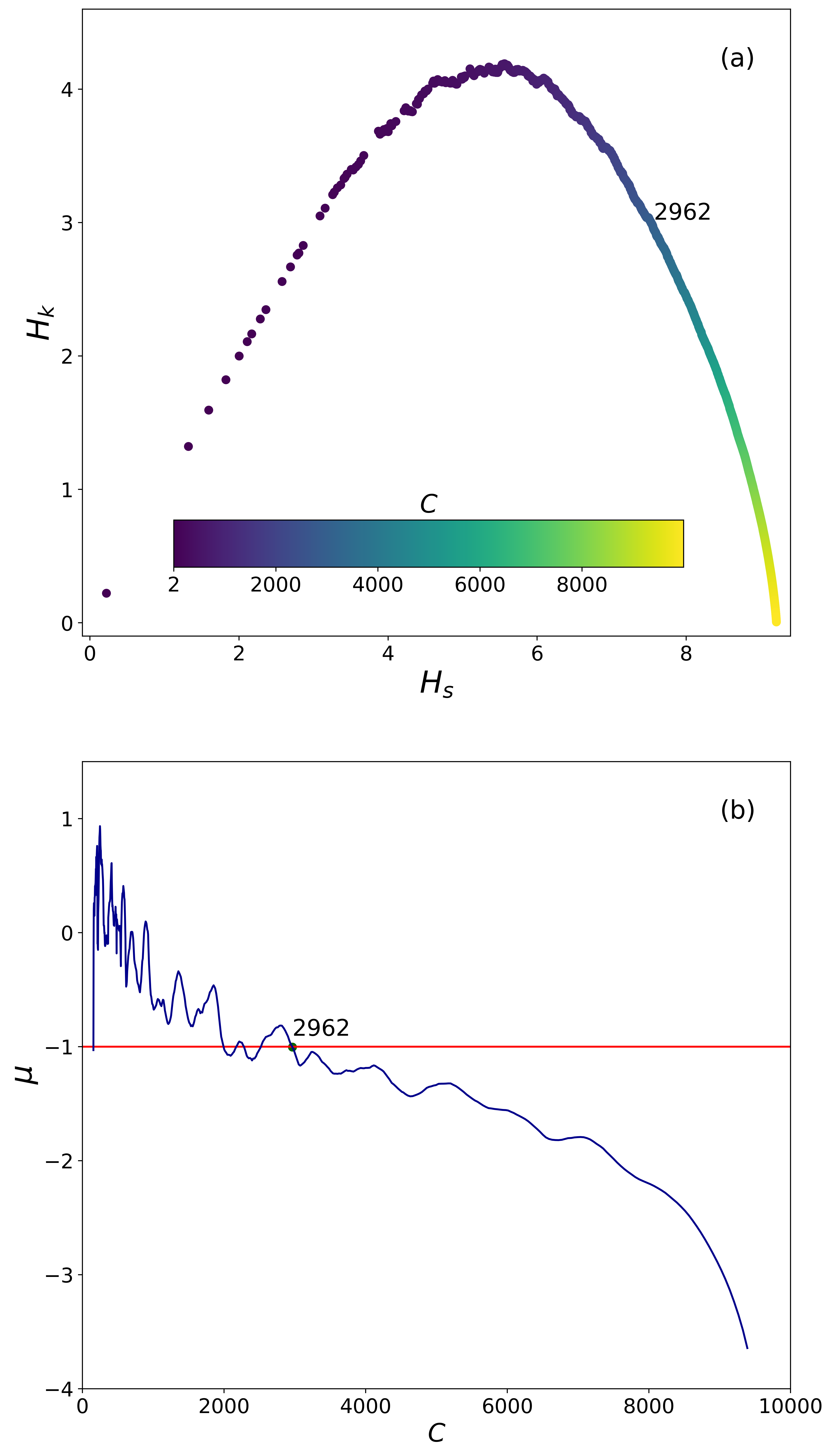}
    \caption{(a): Resolution-relevance plot for 6D93, whose \emph{all-atom} trajectory of $L = 10001$ sampled realisations has been clustered in $2000$ different values of $C$, starting from $C = 2$ and ending with $C = 9997$, with an intermediate step equal to $5$. The UPGMA algorithm \cite{sokal1958statistical} is applied to the fully-atomistic RMSD matrix (Eq.~\ref{eq:rmsd}) to perform the clustering (see Ref.~\cite{giulini2020information}). Each data point is coloured according to the value of $C$ employed to label the original trajectory. (b): local slope $\mu$ of the $H[s]$-$H[k]$ curve over all the spectrum of possible values of $C$. $\mu$ is computed by iteratively performing a linear regression over all values of the curve such that the resolution falls into an interval of amplitude $\frac{\ln L}{50}$. Such resolution window is iteratively moved from right to left by a factor $\frac{\ln L}{1000}$, until points with $H[s] = H[k] \sim 0$ are found.}
    \label{fig:cont_model_6d93}
\end{figure}

Hence, resolution and relevance are employed to determine the optimal number of fine-grained labels, which we denote by $\bar{C}$, used to partition the collected protein structures. In Fig.~\ref{fig:cont_model_6d93}(a) we report the $H[s]$-$H[k]$ dependence for the $10001$ realisations of the system; the considered trajectory displays a flat maximum of the relevance, which remains constant over a wide range of values of $H[s]$ and $C$. The nature of this behaviour is certainly related to the hidden structure of the sample and to the properties of the clustering algorithm used to label its constituent elements.

The separation between the regimes of lossless and lossy compression \cite{cubero2019statistical} operated by the relevance is exploited to select $\bar{C}$. Indeed, $\bar{C}$ is chosen as the value of $C$ corresponding to the critical point where the slope $\mu$ of the resolution-relevance curve is $-1$. The probability of each label $\vec{x}$ is now given by the number of times it is observed in the sample ($k_{{\vec{x}}}/L$, see Eq.~\ref{eq:p_phi}).

The calculation of the optimal $\bar{C}$ that separates the region with $\mu < -1$ from that with $\mu > -1$ is shown in Fig.~\ref{fig:cont_model_6d93}(b). In this context, we choose the first value of $C$ after which $\mu < -1$ for a consistent set of values of $C$, meaning that the induced discretisation remains in the regime of lossy compression for a while. At the end, the original trajectory of $L$ snapshots is converted into its reduced counterpart of $\bar{C}$ protein structures by choosing the first configuration of the sample belonging to each label ${\vec{x}}$. This procedure allows us to determine the reference, empirical probability distribution $p({\vec{x}})$ as the frequency with which each of the $\bar{C}$ sampled high-resolution configurations appears.

\begin{figure}
    \includegraphics[width=\columnwidth]{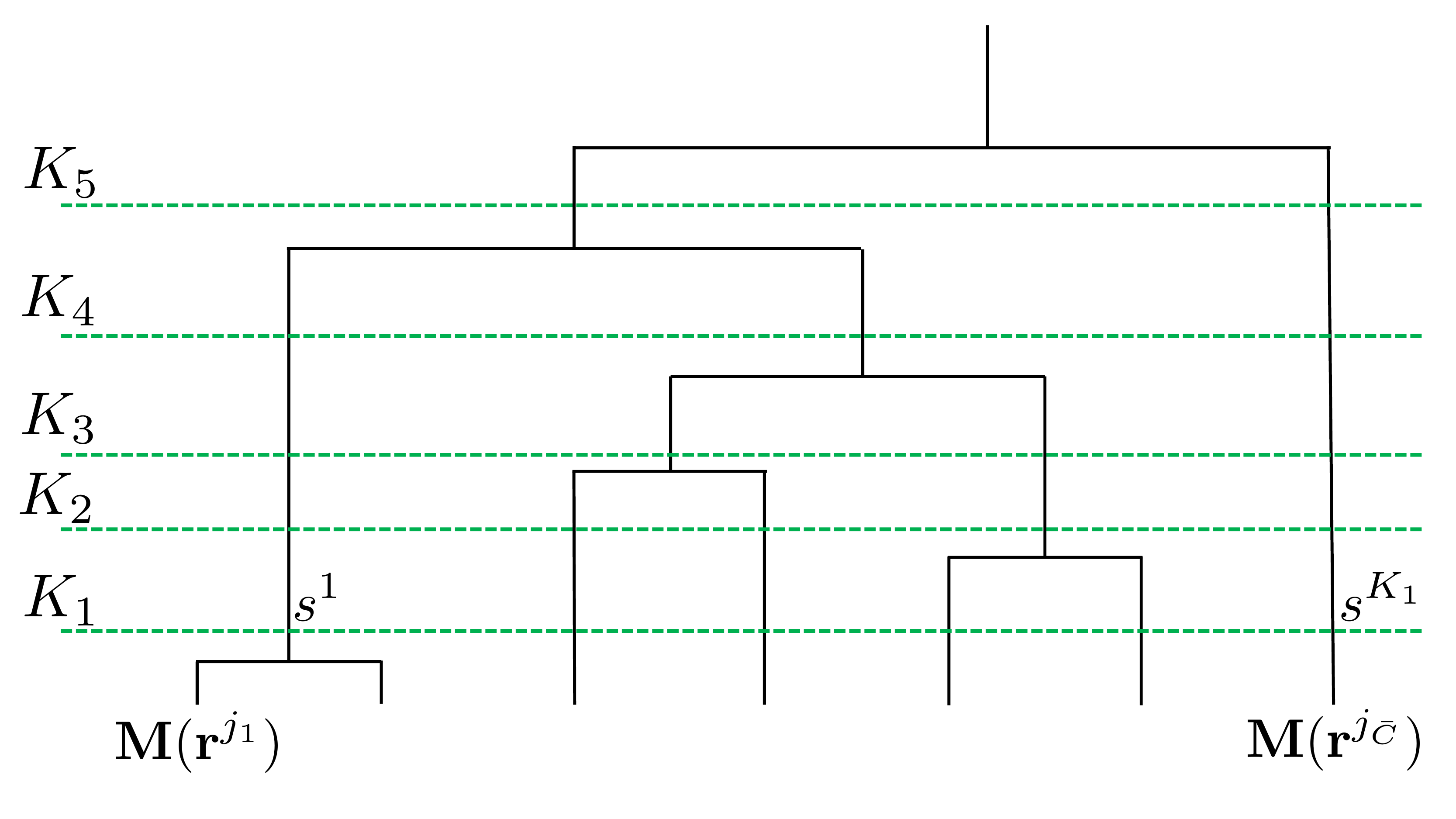}
    \caption{After $\bar{C}$ fine-grained labels are identified (see Fig.~\ref{fig:dendrogram}), the corresponding configurations ${\bf r}^{j_k}$ are coarse-grained by a mapping operator ${\bf M}$. Here $j_k$ is the index (in the original sample) of the configuration associated to the fine-grained label $k$. The low-resolution projections ${\bf M} ({\bf r})$ are first compared and then clustered in a coarse-grained dendrogram. The latter is inspected at several points $K_1$, $K_2$, ... , $K_5$, so as to identify different selections of CG labels $s$, over which the mapping entropy is calculated using Eq.~\ref{eq:ave_smap}. Here we follow Ref.~\cite{giulini2020information} in defining $K_1 = 34$, $K_2 = 48$, $K_3 = 62$, $K_4 = 76$, $K_5 = 91$.}
    \label{fig:cg_dendrogram}
\end{figure}

Next, we consider low-resolution representations of the protein structure, in order to identify the one that provides the most informative picture of the system with respect to the all-atom reference. A CG decimated representation of a protein is a selection of $N$ out $n$ atoms, which amounts at keeping $N$ triplets of the original degrees of freedom. The coarse-grained labelling $s = {\bf M}(r_1, \dots, r_{3n})$ lumps $\bar{C}$ high-resolution labels ${\vec{x}}$ in $K$ CG labels $s$. Following Ref.~\cite{giulini2020information}, we here select $5$ different values of $K$ to cut the dendrogram, thus creating $5$ different probability distributions $\overline{p}({\vec{x}})$. In Fig.~\ref{fig:cg_dendrogram} we provide a schematic depiction of this procedure: first, $\bar{C}$ mapped configurations ${\bf M}(r_{1}, \dots, r_{3n})$, are compared using the \emph{coarse-grained} RMSD
\begin{eqnarray}\label{eq:cg_rmsd}
&&\text{RMSD}^{\text{\tiny{CG}}} ({\bf M}({\bf r}) , {\bf M}({\bf r}^\prime)) =\\ \nonumber
&&= \sqrt{\frac{1}{N} \sum_{j=1}^{N} \sum_{k=1}^{3}(r_{i_{3j+k}} - \mathcal{RT} r^\prime_{i_{3j+k}})^{2}},
\end{eqnarray}
where $i$ denotes the indices of the retained degrees of freedom, and then the corresponding dendrogram is constructed with the UPGMA algorithm. Subsequently, different thresholds are employed to define the CG labels, and the resulting average mapping entropy is calculated using the following formula \cite{giulini2020information}:
\begin{equation}
\label{eq:ave_smap}
\asmap = \frac{1}{|K|}\sum_{\{K\}} \ S_{map}(K),
\end{equation}
where $\{K\}$ is the set of values of $K$ and $S_{map}(K)$ is the corresponding mapping entropy, arising from the clustering of $\bar{C}$ high-resolution labels into $K$ CG labels; $|K| = 5$ is the number of $K$ values.

\begin{figure}[ht]
    \includegraphics[width=\columnwidth]{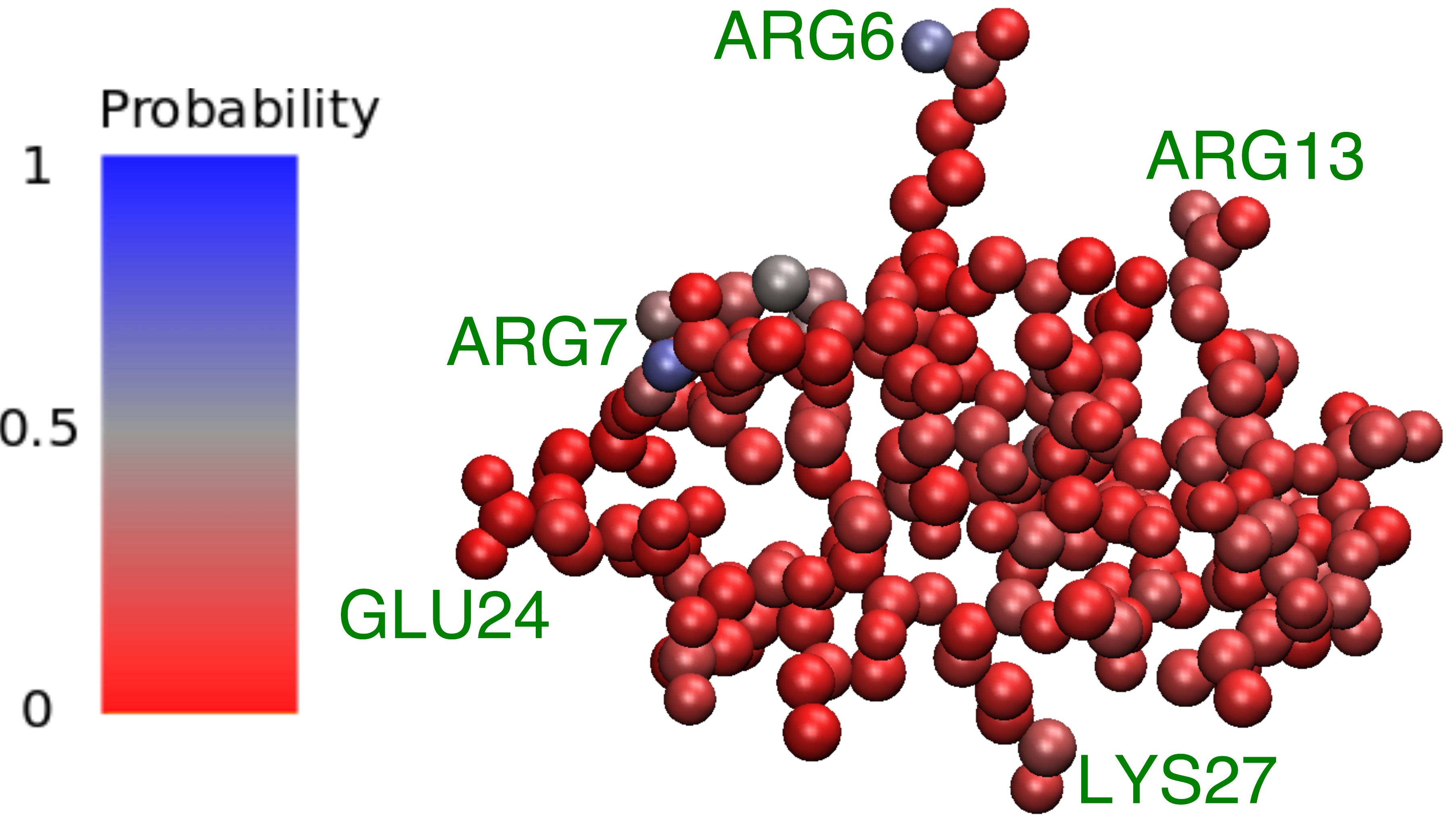}
    \caption{Probability  $P_{cons}$ of conserving each atom in an optimal mapping built minimising \asmap{} (see Eq.~\ref{eq:ave_smap}). Five residues are highlighted, namely the three arginines and other two solvent-exposed, charged residues. While the former are retained with a good level of detail inside optimal mappings (one atom per side chain, see main text), the latter are highly coarse-grained (see also Tab~.\ref{tab:glu24_lys27}).}
    \label{fig:pcons}
\end{figure}

Now that we possess a method to calculate the mapping entropy (Eq.~\ref{eq:smap}) for a continuous system, we follow Ref.~\cite{giulini2020information} and run $48$ mapping optimisations for the protein, employing $N=31$ and sticking to the same minimisation protocol. As in Ref.~\cite{giulini2020information}, one can perform a basic statistics over the pool of low-$S_{map}$ mappings by using the conservation probability, $P_{cons}$, of each atom, defined as the fraction of times it is included inside an optimised solution.

Once projected over the high-resolution protein, such probability distribution appears to be broadly spread throughout the polypeptide chain, with few notable peaks in correspondence of terminal atoms of the three arginine residues of the protein ({\sc ARG6}, {\sc ARG7}, {\sc ARG13}), which are well-known \cite{andreotti2005molecular,quintero2013scorpion,ramirez2014cytotoxicity} to play a crucial role in the binding of tamapin to its substrate. Let us focus on the side chain of {\sc ARG6}: here, the atom with highest \textit{importance} is NH2 ($P_{\text{cons}} (\text{NH2},\text{ARG6})$ = $0.60$), but all the other atoms in the terminal region of the arginine display a non-negligible value of $P_{\text{cons}}$, namely $0.10$, $0.23$, $0.08$ for NE, CZ, and NH1, respectively. The sum of these probabilities with the one associated to NH2 gives $1.02$: except for two (resp. one) cases in which there are two (resp. zero) atoms of this region in the optimal mapping, all the remaining $46$ optimal solutions contain exactly one atom in the terminal region of {\sc ARG6}. In other words, the optimisation procedures are informing the modeller that the side chain of this arginine must be treated with exactly one atom, with a preference for NH2. As for {\sc ARG7} and {\sc ARG13}, they display a similar behaviour, with the majority of the optimisations retaining one atom of their side chain terminus. In particular, the NH2 atom of {\sc ARG7} shows the highest value of $P_{\text{cons}}$ ($P_{\text{cons}} (\text{NH2},\text{ARG7})$ = $0.67$). These results are consistent with those found through the cumulant expansion approximation \cite{giulini2020information}, thus supporting the viability and robustness of this procedure.

In summary, the properties of relevance and resolution are here exploited in order to extract a set of fine-grained labels out of a molecular dynamics trajectory, each one weighted with its own approximated probability. This step is key in order to compute the mapping entropy of a system defined in terms of continuous degrees of freedom: in fact, while for the case of a molecular system in thermal equilibrium approximations are possible, that rely on the assumption of Boltzmann statistics and the cumulant expansion approximation of the mapping entropy (as it was done in Ref. \cite{giulini2020information}), in general the underlying probability density of the system micro-states is not known, and/or it is not an equilibrium distribution. The approach illustrated here is general and unsupervised, and it can be also applied to tasks other than the calculation of the mapping entropy. We deem it important to remark the fact that the choice of distance (RMSD) and clustering algorithm (UPGMA) played no special role in the analysis presented in this section, thus broadening the generality of the proposed approach.

\section{Conclusions}
\label{sec:conclusions}

In this manuscript we investigated the properties of coarse-grained representations by studying the behaviour of the associated resolution, relevance, and mapping entropy, computed over empirical samples of three complex systems. These three quantities offer distinct and complementary perspectives on the properties of a dataset, allowing one to extract crucial information about its underlying generative process, the nonlinear correlations among its degrees of freedom, and the levels of significance of the latter. Mapping entropy, in particular, is employed to characterise a system by quantifying the amount of information retained in a low-dimensional representation of it, which thus highlights those reduced models that preserve relevant details while discarding noise or otherwise trivial features. When coupled, resolution and mapping entropy show, in a clear and easily intelligible way, how the information content of a representation increases together with the detail with which such representation describes the data set.

In the case of the non-interacting spin system, the mapping entropy pinpoints as ideal mappings those subsets of features that match our intuition of most informative representations. In contrast, when the system's constituents are interacting, as it is the case for the model of the Nasdaq stock market, the interpretation of maximally informative coarse-grained mappings is less immediate. Still, the mapping entropy efficiently and consistently separates the stocks that have been influential for the majority of the sampling time from those whose importance has been limited to the last portion of the selected time-window \cite{fernandez2021}. In both cases, the resolution-relevance framework proved to be capable of highlighting the optimal level of detail at which a coarse representation of the system provides the largest amount of nontrivial information about the underlying generative process.

This feature was explicitly employed in addressing the problem of dimensionality reduction for a biomolecule, namely a small protein; in this case, the mapping entropy minimisation requires the knowledge of an underlying, high-resolution probability density that cannot be naively reconstructed from a sample of configurations. To tackle this issue, we proposed a method, based on the optimal trade-off between resolution and relevance, to identify unambiguous high-resolution labels defining a non-uniform probability distribution in the fine-grained space; these corresponds to clusters of configurations whose relative discrepancies are classified as noise by the relevance, thereby allowing the construction of a dataset of high-resolution configurations each associated to its empirical probability. Making use of this protocol, we then carried out several minimisations of the mapping entropy: the resulting optimal representations tend to display an uneven level of detail throughout the protein, treating with higher accuracy the three arginine residues that are fundamental for its binding to the substrate, consistently with data obtained through an independent procedure.

These results, obtained from a relevant set of distinct test cases, show that the combined usage of resolution, relevance, and mapping entropy is capable of quantifying the information content proper to different combinations of features of a high-dimensional, large-sized data set. In particular, it is our opinion that the multi-body nature of the mapping entropy, together with its simplicity of interpretation, can make its application in data science extremely fruitful, either as a feature selection algorithm or as a novel instrument of analysis of complex data sets. The first use is analogous to the mapping definition in CG, that is, a smart prescription to be implemented \textit{prior to the modelling}. The second application is even more intriguing, as it suggests that the process of dimensionality reduction \textit{per se} can provide information on high-dimensional data sets.

Looking at resolution, relevance, and mapping entropy from this multi-disciplinary perspective, it is our opinion that their application in diverse contexts would contribute a powerful instrument to make sense of data in a world increasingly full of them.

\section{Appendix}

\subsection{Explicit derivation of the relation between resolution and mapping entropy}\label{app:1}

Hereafter we provide the full derivation of Eq. \ref{eq:smap_res}, in which each step is made explicit:
\begin{widetext}
\begin{eqnarray}\label{eq:app:1}
S_{map} &=& \sum_{\vec{x}} p(\vec{x}) \ln \frac{p(\vec{x})}{\bar{p}(\vec{x})} \\ \nonumber
&=& \sum_{\vec{x}} p(\vec{x}) \ln \left( p(\vec{x}) \frac{\Omega_1(s(\vec{x}))}{p(s(\vec{x}))} \right)\\ \nonumber
&=& \sum_{\vec{x}} p(\vec{x}) \ln p(\vec{x}) - \sum_{\vec{x}} p(\vec{x}) \ln p(s(\vec{x})) + \sum_{\vec{x}} p(\vec{x}) \ln \Omega_1(s(\vec{x}))\\ \nonumber
&=& - H[\vec{x}] - \sum_{\vec{x}} \sum_s \delta(s(\vec{x}) - s) p(\vec{x}) \ln p(s(\vec{x})) + \sum_{\vec{x}} \sum_s \delta(s(\vec{x}) - s) p(\vec{x}) \ln \Omega_1(s(\vec{x}))\\ \nonumber
&=& - H[\vec{x}] - \sum_s p(s) \ln p(s) + \sum_s p(s) \ln \Omega_1(s)\\ \nonumber
&=& - H[\vec{x}] + H[s] + \sum_s p(s) \ln \Omega_1(s).
\end{eqnarray}
\end{widetext}

As for Eq. \ref{eq:6}, which relates the conditional entropy $H[\vec{x} | s]$ to the conditional probability $p(\vec{x} | s)$, we have:
\begin{eqnarray}\label{eq:app:2}
H[\vec{x} | s] &=& - \sum_{\vec{x}, s} p(\vec{x}, s) \ln \frac{p(\vec{x}, s)}{p(s)}\\ \nonumber
&=& - \sum_{\vec{x}, s} p(\vec{x}, s) \ln p(\vec{x} | s)\\ \nonumber
&=& - \sum_{\vec{x}, s} p(s) p(\vec{x} | s) \ln p(\vec{x} | s)\\ \nonumber
&=& - \sum_{s} p(s) \sum_{\vec{x}} p(\vec{x} | s) \ln p(\vec{x} | s).
\end{eqnarray}

\subsection{Infinite sampling assumption}
\label{sec:infinite-sampling}

Sections \ref{sec:spin} and \ref{sec:nasdaq} discuss the case in which the fine-grained and coarse-grained labels $\vec{x}$ and $s(\vec{x})$ are determined through a marginalisation over the retained degrees of freedom. We now discuss how, in such scenarios, we can apply the assumption of infinite sampling and how it changes the overall results.

\begin{figure}
    \includegraphics[width=\columnwidth]{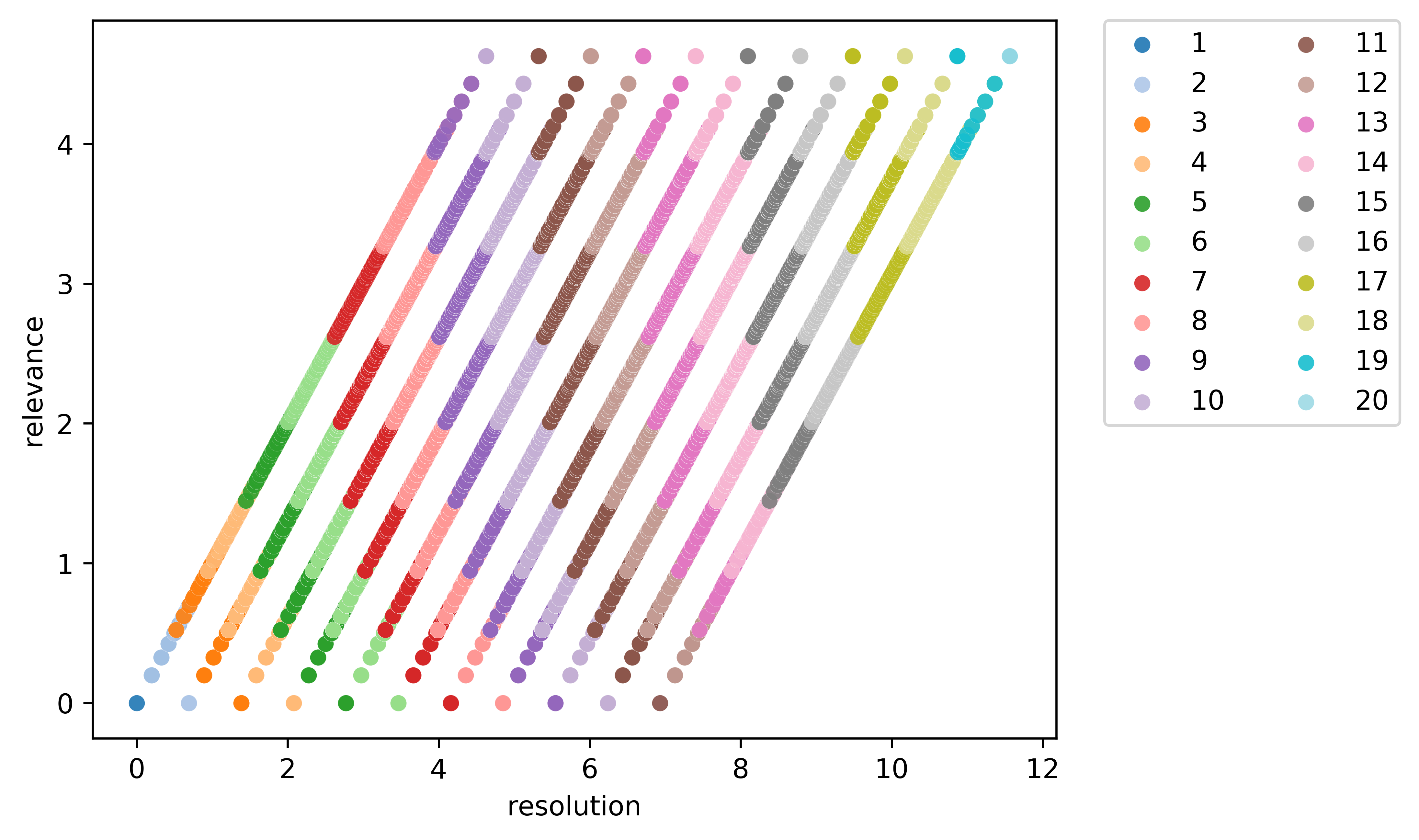}
    \caption{Relevance vs. resolution for the toy  model of binary spins with exhaustive configurational sampling and exact underlying probabilities. The colours refer to the number of retained sites. These results should be compared with the finite sample case in Fig.~\ref{fig:spin}(a).}
    \label{fig:resrel_inf}
\end{figure}

First we investigate the impact on the relevance of having such a large sampling that the empirical probability $\hat{p}(s)$ is arbitrarily close to the real one obtained marginalising over the exact $p(\vec{x})$. To this end, we consider the toy model of Sec.~\ref{sec:spin} and compute resolution and relevance summing over the complete list of all possible states of the system, whose probability is known from Eq.~\ref{eq:2}. In this case, we obtain the resolution-relevance curves reported in Fig.~\ref{fig:resrel_inf}. These are quite different from the empirical ones emerging from a finite sample (see Fig.~\ref{fig:spin}(a)), demonstrating that the relevance shows a non-trivial behaviour even in the case of a simple system.

Let us explain how a finite sample size can create such a qualitatively different behaviour. This happens when the system has some equiprobable configuration $p(s_i) = p(s_j)$ for two distinct CG labels $s_i,\ s_j$. This means that, for an infinite sample, the frequencies of these configurations are equal, i.e. $k_{s_i} = k_{s_j}$.

Let us assume that the frequency $k_{s_i}$ appears exactly twice in the sample, $m_{k_{s_i}} = 2$. This implies $\hat{p}(k_{s_i}) = 2 k_{s_i} / L$. Recall that the relevance is given by summing over the frequencies $k$; therefore, if $k_{s_i} = k_{s_j}$ there is only one term contributing, whereas if $k_{s_i} \neq k_{s_j}$ there are two terms. The contribution of the frequency $k_{s_i}$ to the relevance is
\begin{eqnarray}
  \label{eq:rel-contribution-equiprobable}
  - \hat{p}(k_{s_i}) \log \hat{p}(k_{s_i})  &=& - \frac{2 k_{s_i}}{L} \log \frac{2 k_{s_i}}{L} \\ \nonumber
  &=& - \frac{2 k_{s_i}}{L} \log \frac{k_{s_i}}{L} - \frac{2 k_{s_i}}{L} \log 2.
\end{eqnarray}

When the empirical sample is finite and the frequencies are not equal, $k_{s_i} \neq k_{s_j}$,  but close, $k_{s_i} \approx k_{s_j}$, the contribution to the relevance is
\begin{eqnarray}
  \label{eq:rel-contribution-not-equiprobable}
  && - \hat{p}(k_{s_i}) \log \hat{p}(k_{s_i}) - \hat{p}(k_{s_j}) \log \hat{p}(k_{s_j})   \\
   && = - \frac{k_{s_i}}{L} \log \frac{k_{s_i}}{L} - \frac{k_{s_j}}{L} \log \frac{k_{s_j}}{L} \\
   && \approx - \frac{2 k_{s_i}}{L} \log \frac{k_{s_i}}{L}.
\end{eqnarray}

Comparing these two cases, we observe that the contribution to the relevance of the infinite sample is lower than the finite case by roughly $2 k_{s_i} / L \log 2 = 2 \hat{p}(s_i) \log 2$. Therefore, in case of some equiprobable configurations, observing high values of the relevance relies on finite imperfect sampling; sampling ``too well'' can reduce the relevance substantially.

We now consider the effect of infinite sampling on the mapping entropy. When the configuration of the complex system of interest is sampled for an infinite number of times, the multiplicity of labels $\vec{x}$ mapping onto the same CG label $s$ is given by the analytical degeneracy:
\begin{equation}
\label{eq:omega_inf}
\Omega^{\infty}_1(s)=\sum_{\vec{x}} \ \delta(s(\vec{x}) - s) = V^{n-N},
\end{equation}
where $V$ is the phase space volume accessible to each degree of freedom. Here for simplicity we assume that all degrees of freedom have the same accessible phase space volume.
In this case, the mapping entropy can be expressed as a difference of two Kullback-Leibler divergences, where the probabilities $p(\vec{x})$ and $p(s)$ are compared to the uniform distributions $V^{-n}$ and $V^{-N}$, respectively:
\begin{eqnarray}
\label{eq:conf}
S_{\vec{x}} &=& - \sum_{\vec{x}} \ p({\vec{x}}) \ln \left( V^n p({\vec{x}}) \right), \\
S_{s} &=& - \sum_{s} \ p(s) \ln \left( V^N p(s) \right).
\end{eqnarray}

Here, $S_{\vec{x}}$ and $S_{s}$ quantify the gain in information guaranteed by employing $p(\vec{x})$ and 
$p(s)$ to sample the phase space in place of the uniform probability, respectively. The \quotes{infinite-sampling} mapping entropy can be expressed as a difference between these quantities \cite{rudzinski2011coarse,foley2015impact,giulini2020information,kidder2021energetic}:
\begin{equation}
\label{eq:smap_inf_first}
S^{\infty}_{map} = \sum_{\vec{x}} p(\vec{x}) \ln \left(\frac{p(\vec{x})}{p(s(\vec{x}))} \Omega^{\infty}_1(s(\vec{x})) \right) = S_{s} - S_{\vec{x}},
\end{equation}
which is still a strictly non-negative Kullback-Leibler divergence. As an example, let us focus on the case of the approximate financial market discussed in Sec.~\ref{sec:nasdaq}, where each stock can assume three different values ($V=3$, see Fig.~\ref{fig:nasdaq_ex}). In this case $S^{\infty}_{map}$ reads
\begin{equation}
\label{eq:smap_inf_spin}
S_{map}^{\infty} = (n-N) \ln 3 + H[s] - H[\vec{x}].
\end{equation}
$S_{map}^{\infty}$ can be decomposed in a constant term, proportional to $n-N$, accounting for the inherent loss of information arising from retaining fewer stocks, and a difference of resolutions. As the fine-grained entropy is fixed for all CG mappings, the only expression that varies with the mapping ${\bf M}$ is the coarse-grained resolution, $H[s]$. In order to decrease $S_{map}^{\infty}$, the mapping must induce a CG distribution $p(s)$ with low entropy. It is useful to note that $H[\vec{x}]$ (resp. $H[s]$) cannot exceed $n \ln 3$ (resp. $N \ln 3$), which corresponds to the maximum entropy over the possible $3^n$ fine-grained (resp. $3^N$ coarse-grained) labels.

\begin{figure}
    \includegraphics[width=\columnwidth]{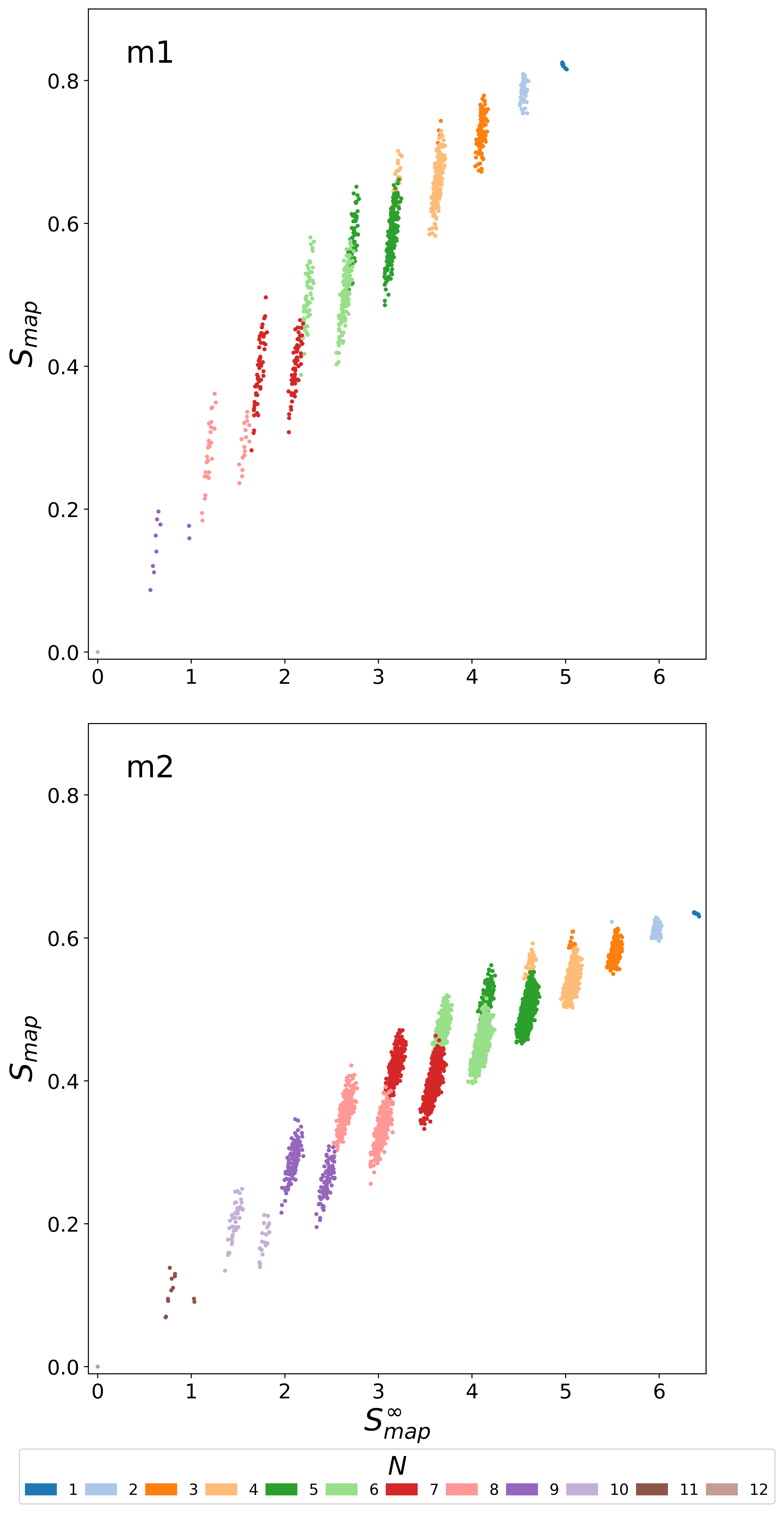}
    \caption{Comparison between the values $S_{map}$ given by Eq.~\ref{eq:smap} and those of $S_{map}^{\infty}$ (Eq.~\ref{eq:smap_inf_spin}). The top and bottom panels correspond to the models \textit{m1, m2} considered in Sec.~\ref{sec:nasdaq}, respectively. Points are coloured according to the number $N$ of retained degrees of freedom, ranging from $1$ to $n$.}
    \label{fig:smap_inf}
\end{figure}

Figure \ref{fig:smap_inf} reports the comparison between the values of $S_{map}^{\infty}$ (Eq.~\ref{eq:smap_inf_spin}) and those of $S_{map}$ (Eq.~\ref{eq:smap}) for the two models \textit{m1, m2}, considered in Sec.~\ref{sec:nasdaq}. Since $S_{map}^{\infty}$ discriminates coarse-grained mappings according to their value of CG resolution $H[s]$, there are two clouds of points for each value of $N$, separating those representations containing both GOOG and GOOGL stocks from the others. However, restricting the analysis to a specific cloud of points, we observe that a distinct, positive correlation exists between the collected values, which is quantified by Tab.~\ref{tab:pearson} in terms of the Pearson correlation and linear coefficients. The correlation is weak when $N=2$, growing up to values higher than $0.8$ for mappings with $N \sim n$. Equation~\ref{eq:smap_res} sheds light on the presence of such correlation, showing that a reduction of the coarse-grained resolution is beneficial for a CG mapping \emph{if and only if} it is not counterbalanced by an increase in $\sum_{\vec{x}} \ p(\vec{x}) \ln \left( \Omega_1 (s(\vec{x}))\right)$. The latter situation is experienced by mappings containing both GOOG and GOOGL stocks, which possess a low-$H[s]$ probability distribution not because of their informativeness, but only because the number of resolved CG labels $s$ is limited.

\begin{table}[htbp]
\centering
  \begin{tabular}{|c|c|c|c|c|c|c|c|c|}
    \hline
    \multicolumn{1}{|c|}{} &\multicolumn{4}{|c|}{m1} & \multicolumn{4}{|c|}{m2}\\
     \cline{2-9}
     %\addlinespace[0.1cm]
    $\quad N\quad$  & $r_{G}$ & $q_{G}$ & $r_{\bar{G}}$ & $q_{\bar{G}}$ & $r_{G}$ & $q_{G}$ & $r_{\bar{G}}$ & $q_{\bar{G}}$ \\\hline
    $\quad2 \quad$& $/$ & $/$ & $0.23$ & $0.16$& $/$ & $/$ & $0.15$ & $0.05$ \\\cline{2-9}    
    $\quad3 \quad$& $\;0.25 \;$ & $\;0.22\;$ & $\;0.54 \;$ & $\;0.47\;$ &$\;0.22\;$ & $\;0.10\;$ &$\;0.43\;$ & $\;0.18\;$ \\\cline{2-9}
    $\quad4 \quad$& $ 0.54 $ & $0.58$ & $0.67$ & $0.62$ & $0.45$ & $0.21$ &$0.55$ &  $0.23$\\\cline{2-9}
    $\quad5 \quad$& $\; 0.69 \;$ & $0.79$ & $0.75$ & $0.70$ & $0.56$ & $0.26$ &$0.62$ &  $0.26$ \\\cline{2-9}
    $\quad6 \quad$& $\; 0.78 \;$ & $0.91$ & $0.79$ & $0.75$ & $0.62$ & $0.28$ &$0.66$ & $0.28$\\\cline{2-9}
    $\quad7 \quad$& $\; 0.84 \;$ & $1.00$ & $0.81$ & $0.77$ & $0.66$ & $0.30$ &$0.70$ & $0.31$\\\cline{2-9}
    $\quad8 \quad$& $\; 0.88 \;$ & $1.06$ & $0.75$ & $0.70$ & $0.71$ & $0.33$ &$0.72$ & $0.34$ \\\cline{2-9}
    $\quad9 \quad$& $\; 0.90 \;$ & $1.03$ & $/$ & $/$ & $0.75$ & $0.37$&$0.73$ & $0.37$ \\\cline{2-9}
    $\quad10\quad$& $/$ & $/$ & $/$ & $/$ & $0.79$ & $0.44$ & $0.72$ & $0.42$ \\\cline{2-9}
    $\quad11\quad$& $/$ & $/$ & $/$ & $/$ & $0.82$ & $0.55$ & $/$ & $/$\\\cline{2-9}
   \hline  
  \end{tabular}
  \caption{Pearson correlation ($r$) and linear interpolation ($q$) coefficients between the values of $S_{map}^{\infty}$ and those of $S_{map}$ (see Fig.~\ref{fig:smap_inf}) for the two models considered. At each value of the number of CG sites $N$, the coefficients are calculated considering mappings containing \emph{both} Google stocks ($r_{G}$, $q_{G}$) and representations in which there is at most one of the two Google stocks ($r_{\bar{G}}$, $q_{\bar{G}}$).}
  \label{tab:pearson}
\end{table}

\subsection{Mapping Entropy of biomolecules: a comparison with previous strategies}

In two recent articles \cite{giulini2020information, errica2021deep}, some of us exploited the mapping entropy as an instrument to explore the space of coarse-grained mappings of 6D93. In order for this quantity to be computed from a fully atomistic MD trajectory, a few approximations are employed in these works, giving rise to the approximated mapping entropy:
\begin{eqnarray}
\label{eq:smap_cum.1_main}
S^{\beta}_{map} &\simeq& k_{B} \frac{\beta^2}{2} \sum_{s} p(s) \langle(U-\langle U\rangle_s)^2\rangle_s,
\end{eqnarray}
where the sum runs over CG labels $s$, $U$ is the potential energy of the system and the subscript $s$ denotes an average conditioned to the CG label $s$. In other words, the approximate mapping entropy of a CG label $s$ is given by the variance of the potential energy proper of those high-resolution configurations $\vec{x}$ mapping onto it.

Such approximations are necessary due to the incapacity of calculating the probability distributions involved in the definition of the mapping entropy, which is the canonical average of the logarithm of $p(\vec{x})/\overline{p}(\vec{x})$ (see Eq.~\ref{eq:smap}). Both $p(\vec{x})$ and $\overline{p}(\vec{x})$ are complicated to extract because of their high dimensionality and the numerical instabilities associated to the explicit calculations of the exponentials.

\begin{figure}
    \includegraphics[width=\columnwidth]{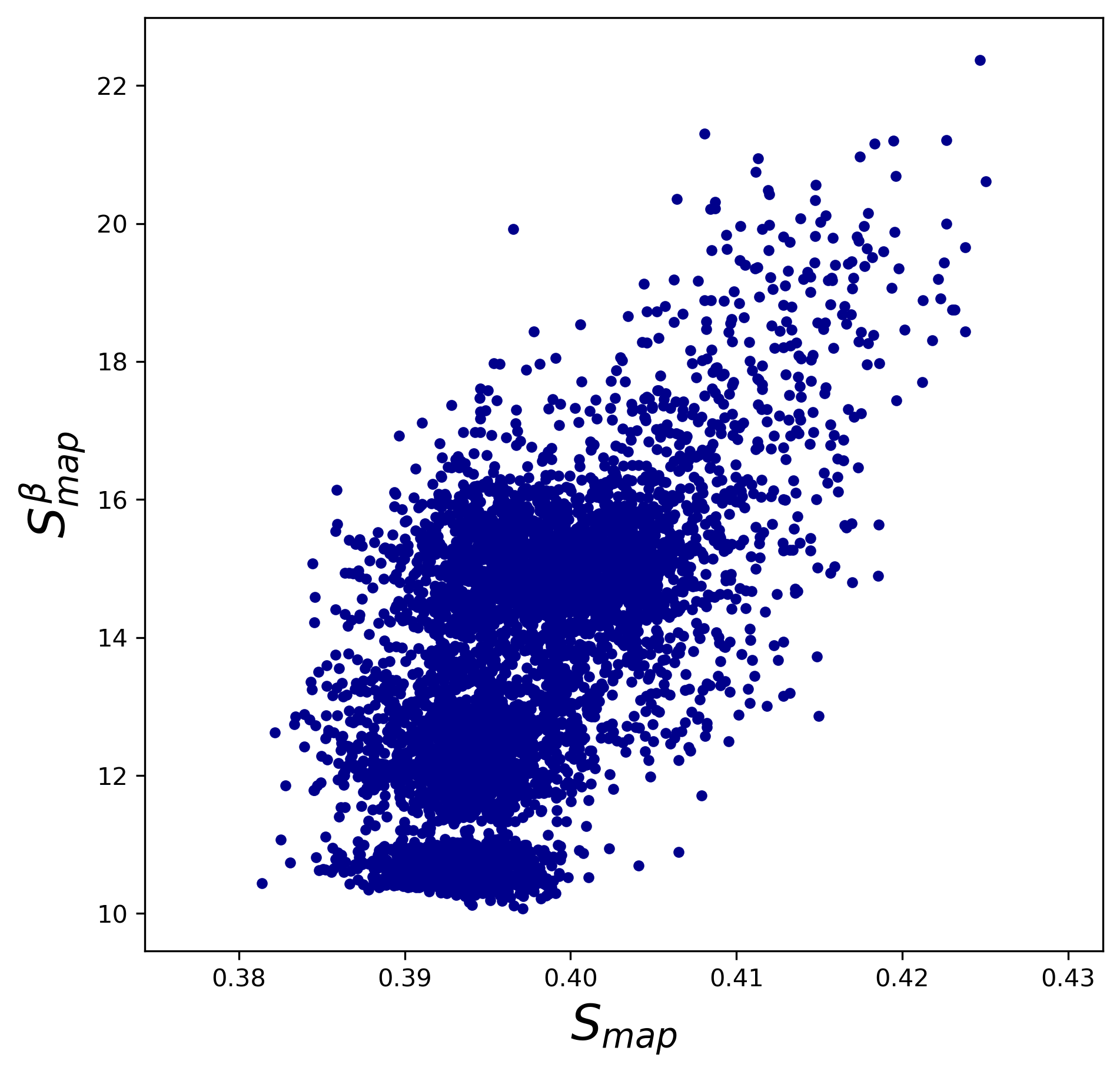}
    \caption{Comparison of the values of mapping entropy calculated using the original Kullback-Leibler formula ($S_{map}$, Eq.~\ref{eq:smap}) and the approximated expression $S^{\beta}_{map}$ of Eq.~\ref{eq:smap_cum.1_main} \cite{giulini2020information}, expressed in units of $kJ/\text{mol}/K$. The protein displays a clear correlation between the two expressions, resulting in a Pearson correlation coefficient equal to $0.62$.}
    \label{fig:smap_comparison}
\end{figure}

Analogously to Eq.~\ref{eq:ave_smap}, we can define an average mapping entropy
\begin{equation}
\label{eq:ave_smap_beta}
\asmap^{\beta} = \frac{1}{|K|}\sum_{\{K\}} \ S_{map}^{\beta}(K).
\end{equation}
In Fig.~\ref{fig:smap_comparison} we report the comparison of the values of \asmap{} and $\asmap^{\beta}$ calculated for the data set of $4968$ mappings with $N = 31$ employed in Ref.~\cite{errica2021deep}. The scatter plot shows that a good but not perfect correspondence exists between the two sets of values. It is important to underline how the nature of the energy considered in the calculation of $S^{\beta}_{map}$ can possibly play a role in this difference: indeed, $S^{\beta}_{map}$ is computed employing only the protein-protein interaction energy, thus neglecting protein-solvent and solvent-solvent effects. Such approximation can give rise to a bias towards exposed regions, where the interactions are not properly screened. One of the strengths of $S_{map}$ is represented by the fact that the solvent contribution is taken into account more accurately by the probability. Overall, further work is needed to assess the nature of this discrepancy.

\begin{figure}[ht]
    \includegraphics[width=\columnwidth]{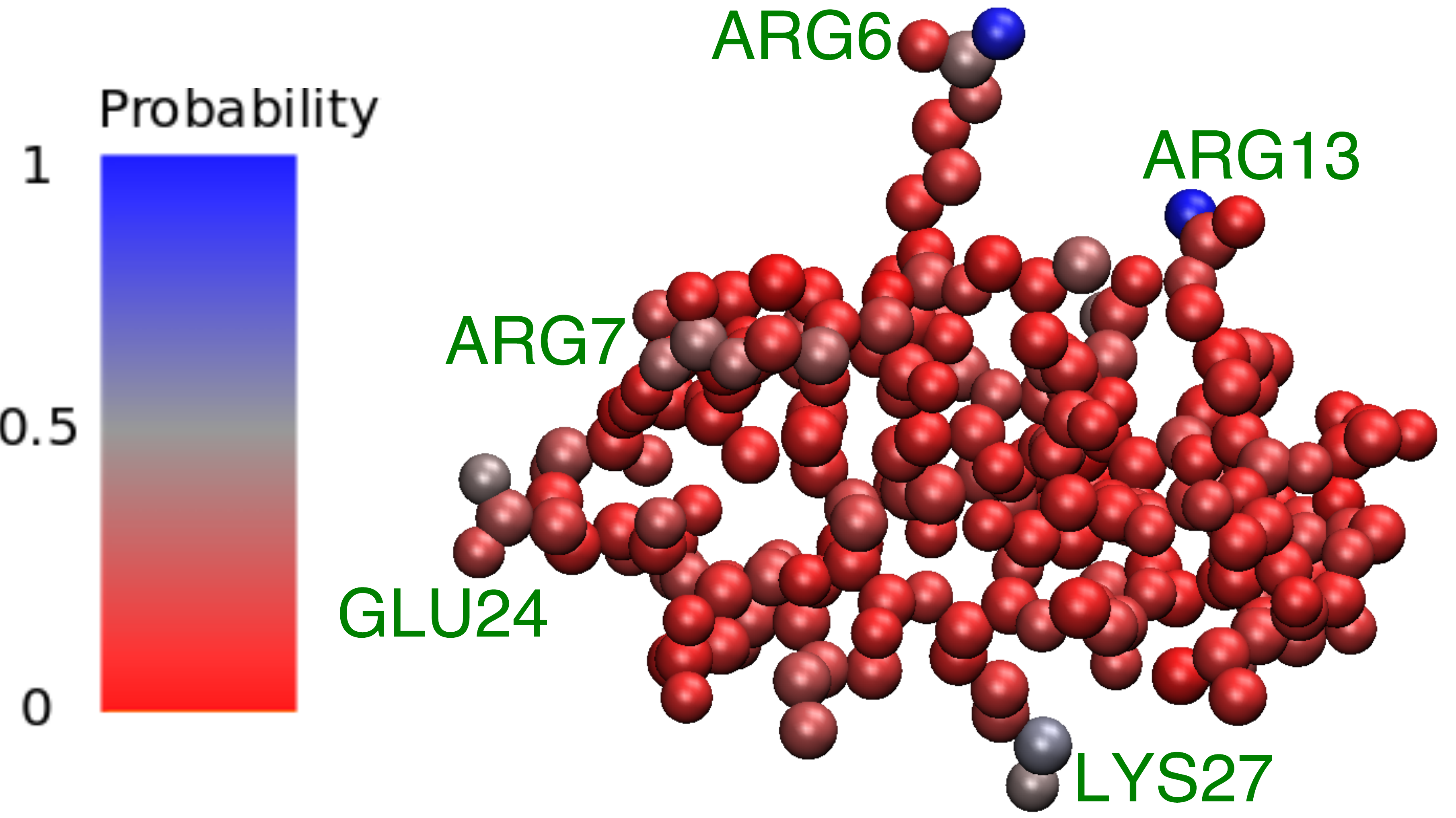}
    \caption{Probability $P_{cons}^{\beta}$ of conserving each atom in an optimal mapping built minimising $\asmap{}^{\beta}$. With respect to $P_{cons}$, $P_{cons}^{\beta}$ is more concentrated in the terminal regions of charged residues, showing more pronounced peaks in correspondence of peculiar atoms.} 
    \label{fig:pcons_beta}
\end{figure}

Analogously to Sec.~\ref{sec:cont} we now analyse the $48$ optimal mappings obtained minimising $\asmap{}^{\beta}$ (see Ref.~\cite{giulini2020information}), with the aim of comparing the resulting conservation probability $P_{cons}^{\beta}$ to the one considered in the main text ($P_{cons}$). Figure \ref{fig:pcons_beta} shows how an optimal CG mapping of 6D93 must contain the NH1 atom ($P^{\beta}_{\text{cons}} (\text{NH1},\text{ARG6})$ = $0.92$), while $P_{cons}$ is more evenly distributed throughout the variable region of this amino acid (see Fig.~\ref{fig:pcons} and Sec.~\ref{sec:cont}). Another interesting difference emerging from a comparison between Fig.~\ref{fig:pcons} and Fig.~\ref{fig:pcons_beta} concerns the reduced values of conservation probabilities assigned to terminal atoms of the variable regions of {\sc GLU24} and {\sc LYS27}; while these atoms were usually part of low-$\asmap{}^{\beta}$ mappings, they are almost never present in the CG representations built minimising $\asmap{}$. {\sc GLU24} and {\sc LYS27} are charged residues, and the energetic fluctuations proper to the terminal atoms can be huge, especially when the considered energies are not screened by the solvent. This is a further proof that $S_{map}$ is less biased towards solvent-exposed, charged residues than $S^{\beta}_{map}$.

\begin{table}
\centering
\begin{tabular}{|P{2.5cm}|P{2.5cm}|P{2.5cm}|}
    \toprule
       \hline  Atom &  $P^{\beta}_{\text{cons}}$ & $P_{\text{cons}}$ \\ \hline 
         %\addlinespace[0.1cm]
    	GLU24-CD  & $0.27$ & $0.00$  \\
        GLU24-OE1  & $0.21$ & $0.00$ \\
        GLU24-OE2 & $0.44$ & $0.02$\\
    	LYS27-CE & $0.52$ & $0.17$\\
     	LYS27-NZ & $0.44$ & $0.27$ \\ \hline
    \end{tabular}
    \caption{Differences between the values of conservation probabilities for the terminal atoms of residues {\sc GLU24} and {\sc LYS27}. The difference is striking especially for {\sc GLU24}, as its terminal atoms are never conserved in the Kullback-Leibler-based optimisation.}
    \label{tab:glu24_lys27}
\end{table}

Overall, it is possible to conclude that Fig.~\ref{fig:pcons} and Fig.~\ref{fig:pcons_beta} are quite similar, with $P_{\text{cons}}$ that is, on average, more evenly distributed over the full structure, displaying a tendency to reduce the probability weight assigned to terminal atoms of charged residues with respect to $P^{\beta}_{\text{cons}}$.

\section{Data availability}

The program and the data employed for the two models presented in Sec.~\ref{sec:spin} and \ref{sec:nasdaq}, as well as the results showed in Sec.~\ref{sec:cont} are available from the GitHub repository at the address \url{https://github.com/mgiulini/pymap} as well as on the Zenodo repository at the address: \url{https://zenodo.org/record/6284439#.YhkWnO7MLUI}.

\section{Acknowledgments}

The authors thank Luca Tubiana, Gianluca Teza and Roberto Menichetti for a critical reading and insightful comments. This project received funding from the European Research Council (ERC) under the European Union's Horizon 2020 research and innovation program (Grant 758588).

\section{Author contributions}

RP conceived the study and proposed the method. RH and MG wrote the softwares, performed the simulations, and collected and analysed the data. All authors contributed to the analysis and interpretation of the data. All authors drafted the paper, reviewed the results, and approved the final version of the manuscript.

\section{Competing interests}

The authors declare no competing interests.

\appendix

\bibliographystyle{ieeetr}
\bibliography{main}

\end{document}